%% file: main.tex
\documentclass[11pt]{article}

\usepackage{geometry}
\usepackage{makecell}
\usepackage{graphicx}
\usepackage{amsmath}
\usepackage{amsthm}
\usepackage{enumerate}
\usepackage{booktabs}
\usepackage{algorithm}
\usepackage{algorithmic}

\usepackage{latexsym}

\usepackage{multirow}
\usepackage{xcolor}
\usepackage{comment}
\usepackage{xspace}

\input{macros}

\newif\iflong
\longtrue 

\newif\ifdraft
\drafttrue 

\newif\ifhideproofs

\ifhideproofs
\usepackage{environ}
\NewEnviron{hide}{}

\fi

\ifdraft
\newcommand{\nb}[1]{\textcolor{red}{\bf!}%
	\marginpar[\parbox{14.5mm}{\raggedleft\scriptsize\textcolor{red}{#1}}]%
	{\parbox{14.5mm}{\raggedright\scriptsize\textcolor{red}{#1}}}}
\newcommand{\warn}[1]{\textcolor{red}{#1}}
\else
\newcommand{\nb}  [1]{}
\newcommand{\warn}[1]{}
\fi

\begin{document}

\title{Consistent Query Answering for Expressive Constraints \\
under Tuple-Deletion Semantics}


\author{
    Lorenzo Marconi
    and
    Riccardo Rosati \\
    Dipartimento di Ingegneria Informatica, Automatica e Gestionale\\
    Sapienza Universit\`a di Roma, Italy}
    
\date{}

\sloppy

\maketitle

\begin{abstract}
We study consistent query answering in relational databases. We consider an expressive class of schema constraints that generalizes both tuple-generating dependencies and equality-generating dependencies. We establish the complexity of consistent query answering and repair checking under tuple-deletion semantics for different fragments of the above constraint language. In particular, we identify new subclasses of constraints in which the above problems are tractable or even first-order rewritable.
\end{abstract}

\newcommand{\A}{\mathcal{A}}
\newcommand{\C}{\mathcal{C}}
\newcommand{\D}{\mathcal{D}}
\newcommand{\E}{\mathcal{E}}
\newcommand{\F}{\mathcal{F}}
\newcommand{\M}{\mathcal{M}}
\newcommand{\N}{\mathcal{N}}
\renewcommand{\O}{\mathcal{O}}
\renewcommand{\P}{\mathcal{P}}
\newcommand{\R}{\mathcal{R}}
\newcommand{\Q}{\mathcal{Q}}
\renewcommand{\S}{\mathcal{S}}
\newcommand{\tup}[1]{\langle #1 \rangle}
\newcommand{\complement}{\textit{Compl}}
\newcommand{\secrets}{\textit{Secrets}}
\newcommand{\olp}{\overline{p}}
\newcommand{\olt}{\overline{t}}
\newcommand{\occurs}{\textit{occurs}}
\newcommand{\atoms}{\textit{Atoms}}
\newcommand{\standardatoms}{\textit{PredAtoms}}
\newcommand{\ineqatoms}{\textit{IneqAtoms}}
\newcommand{\MGU}{\textit{MGU}}
\newcommand{\ol}[1]{\overline{#1}}
\newcommand{\se}{\textrm{ :-- }}
\newcommand{\nnot}{\textit{ not }}
\newcommand{\pred}{\textit{Pred}}
\newcommand{\conj}{\textit{Conj}}
\newcommand{\aux}{\textit{aux}}
\newcommand{\nalpha}{\ol{\alpha}}

\newcommand{\facts}{\textit{Facts}}
\newcommand{\body}{\textit{body}}
\newcommand{\head}{\textit{head}}
\newcommand{\appl}{\textit{Appl}}
\newcommand{\SD}{\tup{\Sigma,\D}}

\input{1-introduction}

\iflong
\input{2-related-work}
\fi
\input{3-preliminaries}
\input{4-repairs}
\input{5-lower-bounds}
\input{6-weak-consistency}

\input{7-repair-checking}
\input{8-query-entailment}

\input{9-conclusions}

\bibliographystyle{abbrv}
\bibliography{bibliography}

\end{document}

%% file: macros.tex
\newtheorem{theorem}{Theorem}
\newtheorem{corollary}{Corollary}
\newtheorem{proposition}{Proposition}
\newtheorem{lemma}{Lemma}

\newtheorem{definitionAux}{Definition}
\newenvironment{definition}{\begin{definitionAux}\rm}{\end{definitionAux}}

\newtheorem{claimAux}{Claim}

\newtheorem{exampleAux}{Example}
\newenvironment{example}{\begin{exampleAux}\rm}{\end{exampleAux}}

\newtheorem{examplesAux}{Examples}

\newtheorem{constructionAux}{Construction}

\newenvironment{proofsk}{\noindent \textsl{Proof (sketch).\ }}{\qedfull}

\long\def\eatpar#1{%
\ifx#1\par                      
\let\nextmove=\eatpar           
\else
\let\nextmove=#1
\fi
\nextmove
}

\def\qed{\hfill{\qedboxempty}      
  \ifdim\lastskip<\medskipamount \removelastskip\penalty55\medskip\fi}

\def\qedboxempty{\vbox{\hrule\hbox{\vrule\kern3pt
                 \vbox{\kern3pt\kern3pt}\kern3pt\vrule}\hrule}}

\def\qedfull{\hfill{\qedboxfull}   
  \ifdim\lastskip<\medskipamount \removelastskip\penalty55\medskip\fi}

\def\qedboxfull{\vrule height 4pt width 4pt depth 0pt}

\newcommand{\aczero}{\mathrm{AC}^0}
\newcommand{\pidue}{\mathrm{\Pi}^p_2}
\newcommand{\logspace}{LOGSPACE}

\newcommand{\constr}{\Sigma}

\newcommand{\fc}{\textit{FC}}
\newcommand{\reps}{\mathsf{rep}_\constr}
\newcommand{\intrep}{\mathsf{intRep}_\constr}

\newcommand{\ars}{AllRep\xspace}
\newcommand{\irs}{IntRep\xspace}

\newcommand{\ra}{\rightarrow}

\newcommand{\wrt}{\iflong with respect to\fi\xspace}

%% file: 1-introduction.tex
\section{Introduction}
\label{sec:introduction}
\newcommand{\tablenote}[1]{\textsuperscript{(#1)}}
\newcommand{\lbcm} {\tablenote{1}}
\newcommand{\ubwi} {\tablenote{2}}
\newcommand{\lbak} {\tablenote{3}}
\newcommand{\ubsc} {\tablenote{4}}
\newcommand{\ubfsh}{\tablenote{5}}
\newcommand{\ubak} {\tablenote{6}}

\newcommand{\comp}{}
\newcommand{\piduecomp}{$\pidue$\comp}
\newcommand{\NP}{NP\comp}
\newcommand{\coNP}{coNP\comp}
\newcommand{\NL}{NL\comp}
\newcommand{\PTIME}{PTIME\comp}
\newcommand{\ffk}{FDET\xspace}
\newcommand{\ded}{DED\xspace}
\newcommand{\deds}{DEDs\xspace}
\newcommand{\dedneq}{DED$^\neq$\xspace}
\newcommand{\inlinesubsection}[1]{\medskip \noindent \textbf{#1}}
\emph{Consistent query answering (CQA)} \cite{ABC99} is the problem of evaluating queries over an inconsistent database that violates its schema constraints.
This problem has been extensively studied in the last years, both in the database and in the knowledge representation area \cite{B19}.

Consistent query answering is based on the notion of \emph{repair} of a database, i.e.\ a modified version of the database that does not violate the schema constraints and is ``as close as possible" to the initial database.
In this paper, we focus on the so-called \emph{tuple-deletion semantics} for database repairs: usually, under such semantics, a repair is a maximal subset of the database that is consistent with the schema constraints.

In the literature on CQA, the schema constraints considered are (subclasses of) \emph{tuple-generating dependencies (TGDs)}, \emph{equality-generating dependencies (EGDs)}, and \emph{universal constraints} \cite{FM05,KW17,B19,W19,KOW21}.
The schema constraints that we consider in this paper are a very large class of database dependencies that captures and generalizes all the above classes of constraints, and correspond to the class of \emph{disjunctive embedded dependencies with inequalities (\deds)} \cite{DT05,D18}.  
A dependency in this class is a first-order sentence corresponding to an implication that uses conjunctions of standard relation atoms and both equality and inequality conditions in the left-hand side of the implication, and a disjunction of conjunctions of the above form in the right-hand side of the implication.

Besides the whole class of \deds, we consider three subclasses: \iflong the class of \fi \emph{linear} \deds, \iflong the class of \fi \emph{acyclic} \deds, and \iflong the class of \fi \emph{forward-deterministic (\ffk)} \deds, as well as all the classes obtained by \iflong pairwisely \fi intersecting these subclasses.

While the first two classes are well-known in the literature, the class of \ffk dependencies is introduced in this paper as a broad generalization of the class of \emph{full} dependencies.

In this paper we study four different decision problems:
\iflong First, we study the complexity of the classical problems of consistent query answering CQA: \fi
\begin{enumerate}[{(1)}]
\item\label{enum-item-rc} \emph{repair checking}, i.e.\ the problem of deciding whether a database $\D'$ is a repair of a database $\D$;
\item\label{enum-item-ar} \emph{(skeptical) query entailment}, i.e.\ the problem of deciding whether a Boolean query is true in every repair of the database. We call this problem \emph{\ars-entailment} in the paper.
\iflong
\end{enumerate}

\noindent
We also study two additional problems:
\begin{enumerate}[{(1)}]
\setcounter{enumi}{2}
\fi
\item\label{enum-item-ir} \emph{intersection-repair query entailment}, i.e.\ the problem of deciding \iflong whether \fi a Boolean query is true in the database corresponding to the intersection of all the repairs of the initial database. We call this problem \emph{\irs-entailment} in the paper;
\item\label{enum-item-wc} \emph{weak consistency}, i.e.\ the problem of deciding whether a database $\D'$ can be extended to a subset of the initial database $\D$ that is consistent with the schema constraints.
\end{enumerate}


The above problems (\ref{enum-item-rc}) and (\ref{enum-item-ar}) are the classical problems of CQA and have been extensively studied in the past: in particular, the most relevant works related to our investigation are \cite{CM05,AK09,SC10,GO10,CFK12,W19}. However, we consider here the larger class of DEDs, the new subclass of FDET dependencies, and a larger query language (\emph{unions of conjunctive queries with inequalities}).
%
Problem (\ref{enum-item-ir}) has been studied recently, mostly in the context of Description Logics ontologies and open-world assumption \cite{LLRRS15,BBG19}.
%
%
As for problem (\ref{enum-item-wc}), we believe that weak consistency is an important problem in the context of CQA. Our analysis is indeed mostly centered around the problem of weak consistency: many complexity results for the other problems follow easily from the complexity of weak consistency, and in many cases both repair checking and query entailment 
can be solved through weak consistency checking techniques.

We study the data complexity of the above problems for the different classes of dependencies obtained combining the acyclicity, linearity and \ffk conditions.
Table~\ref{tab:results-all} presents our results. 

Considering the previous results, the table shows that extending dependencies from TGDs and EGDs to disjunctive embedded dependencies with inequalities and extending the query language from CQs to UCQs with inequalities do not cause any increase in the data complexity of the above reasoning tasks.
Another important aspect shown by the table is the identification of many new tractable cases. Furthermore,
several classes and decision problems are in $\aczero$ (a subclass of \logspace) and can be solved by evaluating a first-order sentence over the database. More precisely:
\begin{itemize}
\item
the \ffk condition has a significant impact on the complexity of all the problems studied, and implies the tractability of both weak consistency and repair checking.
Also, it enjoys the same computational properties of the more restricted class of full TGDs;
\item 
the linearity condition implies the tractability of all the problems studied;
\item
the acyclicity condition has a significant impact on the complexity of all the problems studied, and implies the tractability of repair checking (which is actually in $\aczero$);
\item
the combination of the acyclicity and \ffk conditions implies that all the problems, except \ars-entailment of queries, are in $\aczero$;
\item
the combination of the acyclicity and linearity conditions implies that all the problems are in $\aczero$.
\end{itemize}

We believe that these results are very important not only from the theoretical viewpoint, but also towards the development of practical techniques and tools for consistent query answering.

The paper is structured as follows. 
\iflong
After a brief description of the main related work (Section~\ref{sec:related-work}), in
\fi
Section~\ref{sec:preliminaries} we introduce our languages of dependencies and queries. In Section~\ref{sec:repairs} we introduce the notion of repair and the decision problems studied in the paper, and in Section~\ref{sec:lower-bounds} we show the lower bounds of all the problems. Then, we provide algorithms and prove upper bounds for weak consistency (Section~\ref{sec:wc}), repair checking (Section~\ref{sec:rc}) and both \ars and \irs-entailment (Section~\ref{sec:qe}). \iflong Finally, we \fi conclude in Section~\ref{sec:conclusions}.

\begin{table*}[t]
\centering
\begin{tabular}{|l|c|c|c|c|c|}
\hline
\multirow{2}{*}{\deds} & Weak & Repair & Instance & BUCQ \irs & BUCQ \ars\\
& consistency & checking & checking & Entailment & Entailment \\
\hline
Acyclic $\wedge$ Linear & in $\aczero$ & in $\aczero$ & in $\aczero$ &  \multicolumn{2}{c|}{in $\aczero$} \\
\hline
\ffk $\wedge$ Linear & \NL & \NL & \NL & \multicolumn{2}{c|}{\NL} \\
\hline
Acyclic $\wedge$ \ffk & in $\aczero$ & in $\aczero$ & in $\aczero$ & in $\aczero$ & \coNP~\lbcm \\
\hline

Linear & \PTIME & \PTIME~\ubwi & \PTIME & \multicolumn{2}{c|}{\PTIME~\ubwi} \\
\hline
\ffk & \PTIME & \PTIME~\lbak\ubsc & \coNP & \coNP & \coNP~\lbcm\ubfsh \\
\hline
Acyclic & \NP & in $\aczero$ & \coNP & \coNP & \coNP~\lbcm \\
\hline
All & \NP & \coNP~\lbcm\ubak & \piduecomp~\lbcm & \piduecomp~\lbcm & \piduecomp~\lbcm\ubwi \\
\hline
\multicolumn{6}{l}{\small\makecell[l]{ \\[-2mm]
    \lbcm Lower bound from~\cite{CM05};\\
    \ubwi Extends the upper bound proved in~\cite{CFK12} for the case without inequalities;\\
    \lbak Lower bound from~\cite{AK09};\\
    \ubsc Extends the upper bound proved in~\cite{SC10} for denials;\\
    \ubfsh Extends the upper bound proved in~\cite{CFK12} for GAV (i.e., full single-head) TGDs;\\
    \ubak Upper bound from~\cite{AK09}.
    \smallskip 
}}
\end{tabular}
\caption{Data complexity results. The problems are always complete for the indicated complexity class, except where explicitly stated otherwise. Joined cells indicate that the two entailment problems coincide 
(see Proposition~\ref{pro:cq-entailment-correspondence}).}
\label{tab:results-all}
\end{table*}


%% file: 2-related-work.tex
\section{Related work}
\label{sec:related-work}

\iflong
Consistent Query Answering was originally proposed for relational databases in \cite{ABC99}, which introduced the notions of repairs and consistent answers.
Ever since, many works studied the complexity of CQA considering different kinds of integrity constraints and adopting different repairing approaches.
In particular, as said the introduction, \iflong in the following \fi we consider repairs based on tuple-deletion, with a specific focus on the problems of repair checking and (conjunctive) query entailment.
\fi

For what concerns repair checking, the most relevant works related to our investigation are \cite{CM05,AK09,SC10,GO10,CFK12}, which deeply studied the problem for many classes of dependencies.

Also the problem of finding consistent answers for a given query under tuple-deletion semantics has been extensively studied by \cite{CM05} and \cite{CFK12}, with a particular focus on (some classes of) tuple-generating dependencies, providing complexity results ranging from PTIME to undecidability.
Special attention should also be paid to \cite{FM05} and \cite{KW17}, which proposed first-order rewritable techniques for solving the CQA problem, though limiting the set of integrity constraints to (primary) key dependencies and making some assumptions on the user query.

In the mentioned works, all of which best reviewed in \cite{B19}, the most common semantics adopted corresponds to the one we will call \ars.
The \irs\ semantics, instead, was previously studied in the context of ontologies by \cite{LLRRS10} with the name of IAR (``intersection of ABox repairs"), and further investigated for multiple DL languages in \cite{LLRRS11,R11,LLRRS15}.
In particular, the latter work proved conjunctive query entailment under IAR semantics to be first-order rewritable for the language $DL\text{-}Lite_{R,den}$.
Afterward, \cite{B12} proposed a new semantics named ICR (``intersection of closed repairs"), which IAR is a sound approximation of, showing its first-order expressibility for simple ontologies.
However, such results are not straightforwardly transposable under closed-world assumption, which is our case study.


%% file: 3-preliminaries.tex
\iflong
\section{Preliminaries}
\label{sec:preliminaries}
\subsection{Databases, Dependencies and Queries}
\fi

\newcommand{\ineq}{\gamma}
\newcommand{\equal}{\textit{Eq}}
\newcommand{\notequal}{\textit{NotEq}}
\newcommand{\checkffk}{\textit{Check\ffk}}
\newcommand{\bcq}{\textit{BCQ}}
\newcommand{\cq}{\textit{CQ}}
\newcommand{\vars}{\textit{Vars}}
\newcommand{\varsp}{\textit{PVars}}
\newcommand{\predatoms}{\textit{PA}}
\newcommand{\freevars}{\textit{FVars}}
\newcommand{\cnj}{\textit{Cnj}}
\newcommand{\inequalities}{\textit{Ineq}}
\newcommand{\bodyp}{\predatoms_b}

\iflong
\inlinesubsection{Syntax}
\fi
A \emph{DB schema} $\S$ is a predicate signature, i.e., a set of predicate symbols with an associated arity.
A \emph{term} is either a variable symbol or a constant symbol.
A \emph{predicate atom} (or simply \emph{atom}) is an expression of the form $p(t)$, where $p$ is a predicate of arity $n$ and $t$ is a $n$-tuple of terms. We say that an atom $\alpha$ is ground (or that $\alpha$ is a fact) if there are no occurrences of variables in $\alpha$.
An \emph{inequality atom} (or just \emph{inequality}) is an expression of the form $t\neq t'$, where $t$ and $t'$ are terms.

Given a DB schema $\S$, a \emph{DB instance} (or simply \emph{database}) $\D$ for $\S$ is a set of facts over the predicates of $\S$.
W.l.o.g.\ we assume that a DB schema always contains the special predicate $\bot$ of arity 0, and that no database contains the fact $\bot$.

A \emph{conjunction with inequalities} $\conj$ is a conjunction of predicate atoms and inequality atoms of the form
\begin{equation}
\label{eqn:conj}
\conj=\alpha_1\wedge\ldots\wedge\alpha_k\wedge\ineq_1\wedge\ldots\wedge\ineq_h \end{equation}
where $k\geq 0$, $h\geq 0$, every $\alpha_i$ is a predicate atom and every $\ineq_i$ is an inequality.
We denote by $\vars(\conj)$ the variables occurring in $\conj$, and by $\varsp(\conj)$ the variables occurring in the predicate atoms of $\conj$.
We also denote by $\predatoms(\conj)$ the set of predicate atoms occurring in $\conj$, and by $\inequalities(\conj)$ the subformula $\ineq_1\wedge\ldots\wedge\ineq_h$.

\begin{definition}[CQ and BCQ]
\label{def:cq}
A \emph{conjunctive query with inequalities (CQ)} $q$ is a first-order (FO) formula of the form $\exists y\, (\conj)$, where $\conj$ is a conjunction with inequalities and $y$ is a sequence of variables such that 
$y\subseteq\varsp(\conj)$.\footnote{With a slight abuse of notation, sometimes we use sequences of variables as sets.} We denote by $\freevars(q)$ the set of free variables of $q$ (i.e.\ the variables of $q$ that do not appear in $y$), and define $\vars(q)=\vars(\conj)$, $\varsp(q)=\varsp(\conj)$, $\predatoms(q)=\predatoms(\conj)$ and $\cnj(q)=\conj$. 
When $\vars(q)=\varsp(q)$, we say that $q$ is a \emph{safe} CQ.
When $\freevars(q)=\emptyset$, we say that $q$ is a \emph{Boolean CQ (BCQ)}.
\end{definition}

\begin{definition}[UCQ and BUCQ]
\label{def:ucq}
A \emph{union of conjunctive queries with inequalities (UCQ)} $Q$ is an FO formula of the form
$\bigvee_{i=1}^m q_i$,
where $m\geq 1$ and every $q_i$ is a CQ. We denote by $\cq(Q)$ the set of CQs occurring in $Q$, i.e.\ $\cq(Q)=\{q_1,\ldots,q_m\}$, and denote by $\freevars(Q)$ the set $\bigcup_{q\in\cq(Q)}\freevars(q)$. 
When $\freevars(Q)=\emptyset$, we say that $Q$ is a \emph{Boolean UCQ (BUCQ)}. 
When every $q\in\cq(Q)$ is safe, we say that $Q$ is a \emph{safe} UCQ.
We also define $\vars(Q)=\bigcup_{q\in\cq(Q)}\vars(q)$, $\varsp(Q)=\bigcup_{q\in\cq(Q)}\varsp(q)$, and $\predatoms(Q)=\bigcup_{q\in\cq(Q)}\predatoms(q)$.
\end{definition}

We are now ready to define the notion of dependency that we will use throughout the paper.\footnote{Although slightly different, Definition~\ref{def:dependency-new} is actually equivalent to the notion of disjunctive embedded dependency with inequalities \emph{and equalities} presented in \cite{D18,DT05}.}

\begin{definition}[Dependency]
\label{def:dependency-new}
    Given a DB schema $\S$, a \emph{disjunctive embedded dependency with inequalities} (or simply \emph{dependency}) for $\S$ is an FO sentence over the predicates of $\S$ of the form:
    \begin{equation}
    \label{eqn:dependency}
    \forall x\, \big(\conj \rightarrow Q \big)
    \end{equation}
    where
    $\conj$ is a conjunction with inequalities such that $|\predatoms(\conj)|\geq 1$, $x$ is a sequence of variables such that $\vars(\conj)=\varsp(\conj)=x$,
    and $Q$ is a UCQ such that $\freevars(Q)\subseteq x$.
\end{definition}

Given a DB schema $\S$ and a dependency $\tau$ for $\S$ of the above form (\ref{eqn:dependency}), we indicate with $\body(\tau)$ the formula 
$\conj$, with $\head(\tau)$ the UCQ $Q$, and with $\bodyp(\tau)$ the set $\predatoms(\conj)$.
%
Moreover, we say that $\tau$ is \emph{non-disjunctive} if $m=1$, i.e.\ $\head(\tau)$ is a CQ $q$, and that $\tau$ is \emph{single-head} if it is non-disjunctive and $|\predatoms(\head(\tau))|=1$.


From now on, we will omit that databases and dependencies are given for a DB schema.

\iflong
\inlinesubsection{Semantics}
\fi
Dependencies and safe BUCQs are subclasses of the class of domain-independent relational calculus formulas \cite{AHV95}. 
Given a domain-independent sentence $\phi$ of the relational calculus and a set of facts $\D$ over the same signature, we write $\D\models\phi$ if the evaluation of $\phi$ over $\D$ is true.

Given a database $\D$ and a dependency $\tau$, we sat that $\tau$ is \emph{satisfied in $\D$} iff $\D\models\tau$.
Given a database $\D$ and a set of dependencies $\constr$, we say that $\D$ is \emph{consistent with $\constr$} if all the dependencies of $\constr$ are satisfied in $\D$, i.e.\ if $\D\models\bigwedge_{\tau\in\Sigma}\tau$.

Let $\conj$ be a conjunction with inequalities of the form (\ref{eqn:conj}) and $\D$ be a set of facts over the same signature of $\conj$.
An \emph{instantiation of $\conj$ in $\D$} is a substitution $\sigma$ of the variables occurring in $\conj$ with constants such that 
$(i)$ $\sigma(\alpha_i)\in\D$ for every $i$ such that $1\leq i\leq k$;
$(ii)$ no inequality of the form $t\neq t$ (where $t$ is either a constant or a variable) occurs in  $\inequalities(\conj)$.
Moreover, given an instantiation $\sigma$ of $\conj$ in $\D$, we call \emph{image of $\conj$ in $\D$} the subset 
$\{\sigma(\alpha)\mid\alpha\in\predatoms(\conj)\}$ of $\D$.
%
Given a CQ $q$, an image of $q$ in $\D$ is any image of $\cnj(q)$ in $\D$.
Given a UCQ $Q$, an image of $Q$ in $\D$ is any image of $q$ in $\D$, where $q\in\cq(Q)$.

\iflong
The following property is immediate to verify.
\fi

\begin{proposition}
\label{pro:consistency}
Given a database $\D$ and a safe BUCQ $q$, we have that $\D\models Q$ iff there exists an image of $Q$ in $\D$.
Moreover, a dependency $\tau$ is satisfied by $\D$ iff, for every instantiation $\sigma$ of $\body(\tau)$ in $\D$, there exists an image of the BUCQ $\head(\sigma(\tau))$ in $\D$.
\end{proposition}

We recall that the images of a BCQ with $k$ predicate atoms in a set of facts $\D$ are at most $n^k$ (where $n$ is the cardinality of $\D$) and can be computed in polynomial time \wrt data complexity.

\iflong
\subsection{Subclasses of dependencies}
\fi
We say that a dependency of the form (\ref{eqn:dependency}) is \emph{linear} in the case when $|\predatoms(\conj)|=1$.
Moreover, when $\vars(Q)=\freevars(Q)$, we say that the dependency is \emph{full}.


Given a database $\D$ and a dependency $\tau$, we say that $\tau$ is \emph{forward-deterministic (\ffk) for $\D$} if, for every instantiation $\sigma$ of $\body(\tau)$ in $\D$, there exists at most one image of the BUCQ $\head(\sigma(\tau))$ in $\D$.
Moreover, we say that a set of dependencies $\constr$ is \emph{\ffk for $\D$} if every dependency of $\Sigma$ is \ffk for $\D$.

It is straightforward to see that non-disjunctive full dependencies are forward-deterministic independently of the database $\D$. For disjunctive or non-full set of dependencies $\Sigma$, deciding whether $\Sigma$ is \ffk for a database $\D$ is not a hard task.

\begin{proposition}
\label{pro:ffk-check}
Deciding whether $\Sigma$ is \ffk for $\D$ is $\aczero$ \wrt data complexity.
\end{proposition}
\iflong 
\begin{proof}
First, observe that $\Sigma$ is not \ffk for $\D$ iff there exists a dependency $\tau\in\Sigma$ and an instantiation $\sigma$ of $\body(\tau)$ in $\D$ such that there exists at least two distinct images of $\head(\sigma(\tau))$ in $\D$.
Now let $\checkffk(\constr)$ be the following FO sentence:
\[
\begin{array}{l}
\displaystyle
\bigwedge_{\tau\in\Sigma} \forall x\, \Big(\body(\tau) \ra 
\bigwedge_{q\in\cq(\head(\tau))}\forall y\,\big(\cnj(q)\ra\forall y'\,(\cnj'(q)\ra\equal(y,y'))\big)\Big)
\end{array}
\]
where:
$x$ contains all the universally quantified variables of $\tau$;
$y$ contains all the existentially quantified variables of $q$ (i.e.\ the variables of $q$ not appearing in $\body(\tau)$);
$y'$ contains a copy of all the existentially quantified variables of $q$, and $\cnj'(q)$ is obtained from $\cnj(q)$ by replacing every variable $y_i$ from $y$ with the corresponding copy variable  $y_i'$;
$\equal(y,y')$ is the formula
\[
y_1=y_1' \wedge y_2=y_2' \wedge\ldots\wedge y_m=y_m'
\]
Now, it is immediate to verify that $\D\models\checkffk(\constr)$ iff $\constr$ is \ffk for $\D$. Consequently, the thesis follows. 
\end{proof}
\fi 

Given a set of dependencies $\constr$, we call \emph{dependency graph} of $\constr$ the directed graph $G(\constr)$ whose vertices are the dependencies of $\constr$ and such that there is one edge from the vertex $\tau_1$ to $\tau_2$ iff the head of $\tau_1$ contains an atom whose predicate appears in a predicate atom of the body of $\tau_2$. We say that $\constr$ is \emph{acyclic} if there is no cyclic path in $G(\constr)$.
Given an acyclic set of dependencies $\constr$, we call \emph{topological order of $\constr$} any topological order of $G(\constr)$, i.e.\ a sequence $\tup{\tau_1,\ldots,\tau_h}$ of the dependencies of $\constr$ such that, if $i\geq j$, then the vertex $\tau_i$ is not reachable from the vertex $\tau_j$ in $G(\constr)$.




\iflong
\subsection{Complexity classes}

In this article we refer to the following computational complexity classes:
\begin{itemize}
    \item $\aczero$, i.e. the class of decision problems solvable by a logspace-uniform family of circuits $\{C_n\}$, where $C_n$ has size $O(n^c)$ for some constant $c>0$ and depth $O(1)$, and gates are allowed to have unbounded fan-in;
    \item NL (or NLOGSPACE), i.e. the class of decision problems solvable in logarithmic space by a non-deterministic Turing machine;
    \item PTIME, i.e. the class of decision problems solvable in polynomial time by a deterministic Turing machine;
    \item NP, i.e. the class of decision problems solvable in polynomial time by a non-deterministic Turing machine;
    \item coNP, i.e. the class of decision problems whose complement is solvable in polynomial time by a non-deterministic Turing machine;
    \item $\pidue$, i.e. the class of decision problems whose complement is solvable in polynomial time by a non-deterministic Turing machine augmented by an oracle for some NP-complete problem.
\end{itemize}
These classes are such that $\aczero\subset\text{NL}\subseteq\text{PTIME}\subseteq\text{NP}\cup\text{coNP}\subseteq\pidue$. 
Given a complexity class $C$ and a decision problem $p$, we say that $p$ is \emph{$C$-hard} if there exists a logspace reduction from every other problem in $C$ to $p$. Moreover, if $p$ does also belong to $C$, we say that $p$ is \emph{$C$-complete}.
\fi

%% file: 4-repairs.tex
\section{Repairs and Decision Problems}
\label{sec:repairs}

\newcommand{\variables}{\textit{HVars}}
\newcommand{\equalities}{\textit{Eqs}}

We are now ready to define the notion of repair, the different entailment semantics and the decision problems that we study.

\begin{definition}[Repairs]\label{def:repair}
    Given a database $\D$ and a set of dependencies $\constr$, a \emph{repair} of $\SD$ is a maximal subset of $\D$ that is consistent with $\constr\cup\D$.
\end{definition}

Let us call $\reps(\D)$ the set of all the possible repairs of $\SD$ and let $\intrep(\D)=\bigcap_{D'\in \reps(\D)} D'$.
It is immediate to see that a repair always exists for every $\SD$ (since the empty database satisfies every dependency).

The first decision problems we define are related to two distinct notions of entailment of formulas \wrt the repairs of a database.

\begin{definition}[Entailment]\label{def:entailment}
    Given a database $\D$, a set of dependencies $\constr$, and a domain-independent first-order sentence $\phi$, we say that:
$(i)$ \emph{$\SD$ entails $\phi$ under the \ars\ semantics} (or \emph{$\SD$ \ars-entails $\phi$} for short) if $\D' \models \phi$ for every $\D'\in \reps(\D)$;
$(ii)$ \emph{$\SD$ entails $\phi$ under the \irs\ semantics} (or \emph{$\SD$ \irs-entails $\phi$} for short) if $\intrep(\D) \models \phi$.
\end{definition}

\iflong 
The following example shows that, in general, the two semantics are different.
\begin{example}\label{ex:semantics-difference}
    Let
    $\constr=\{
        \forall x,y,z\, (P(x,y) \wedge P(x,z) \wedge y \ne z \ra \bot),
        \forall x\, (T(x) \ra \exists y\, (P(y,x)))
    \}$ and 
    $\D=\{ P(c,a), P(c,b), P(d,c), \allowbreak T(a), T(b)\}$.
    It's easy to see that the first dependency of $\constr$ is not satisfied by $\D$.
    There are two minimal ways of solving the inconsistency between $\D$ and $\constr$ by means of tuple-deletion. The first one consists in deleting the fact $P(c,a)$ and, due to the second dependency, also the fact $T(a)$. Analogously, the second way consists in deleting both $P(c,b)$ and $T(b)$.
    Thus,
    $\reps(\D)=\{ \{P(c,a), P(d,c), T(a)\}, \{P(c,b), P(d,c), T(b)\} \}$ and
    $\intrep(\D)=\{ P(d,c) \}$.
    Let us now consider the sentence $\phi=\exists x\, (P(c,x))$.
    We have that $\phi$ is entailed by all the repairs, but not by their intersection, i.e.,  is \ars-entailed but not \irs-entailed by $\SD$.
\end{example}
\fi

In the rest of the paper, we will study \emph{UCQ entailment}, i.e.\ the entailment of safe BUCQs, under both semantics. We call will call this problem \emph{instance checking} in the case when the BUCQ is a fact.

The following property, whose proof is immediate, show two important correspondences between the \ars and the \irs semantics of entailment.

\begin{proposition}
\label{pro:instance-checking-correspondence}
Instance checking under the \ars\ semantics coincides with instance checking under the \irs\ semantics.
\iflong
\end{proposition}

\begin{proposition}
\fi
\label{pro:cq-entailment-correspondence}
Moreover, in the case of linear dependencies, entailment under the \ars\ semantics coincides with entailment under the \irs\ semantics.
\end{proposition}

We now define another important decision problem studied in the context of CQA.
Given a database $\D$, a set of dependencies $\constr$, and a set of facts $\D'\subseteq\D$, we call \emph{repair checking} the problem of checking if a $\D'$ is a repair of $\SD$

Finally, we introduce a further decision problem, called \emph{weak consistency}, that is related to the notion of consistency of a database.
\begin{definition}[Weak consistency]
Given a set of dependencies $\constr$, we say that a subset $\D'$ of $\D$ is \emph{weakly consistent with $\SD$} if there exists a subset $\D''$ of $\D$ such that $\D'\subseteq\D''$ and $\D''$ is consistent with $\constr$.
\end{definition}

From now on, in all the decision problems we study, we assume w.l.o.g.\ that all the predicates occurring in $\D$ also occur in $\Sigma$. Moreover, we denote by $\pred(\Sigma)$ the set of predicates occurring in $\Sigma$.

%% file: 5-lower-bounds.tex
\section{Lower bounds}
\label{sec:lower-bounds}

\renewcommand{\succ}{\mathit{Succ}}
\renewcommand{\vert}{\mathit{Vert}}
\newcommand{\propvar}{\mathit{Var}}
\newcommand{\ntv}{\mathit{neg}}
\newcommand{\false}{\textit{false}}
\newcommand{\true}{\mathit{true}}
\newcommand{\initial}{\mathit{Start}}

In this section we provide lower bounds for the decision problems studied in this paper.
We remark that all the lower bounds shown in this section also hold if the dependencies are restricted to be single-head and without inequalities in the head.

\iflong
\subsection{Weak consistency}
\fi
\label{subsec:lb-wc}
We show that the problem of checking weak consistency is NL-hard if the dependencies are both linear and \ffk for the given database. We then show that, if only one of the two properties (i.e., linear or \ffk) is enjoyed, the problem becomes PTIME-hard. Finally, we prove NP-hardness for the case of acyclic dependencies. 

%

\begin{theorem}
\label{thm:wc-lb-linear-ffk}
Weak consistency in the case of linear \ffk dependencies is NL-hard \wrt data complexity.
\end{theorem}
\iflong 
\begin{proof}
We prove the thesis by showing a reduction from STCON (the reachability problem on directed graphs).
Let $\Sigma$ be the following set of linear dependencies:
\[
\begin{array}{l}
\forall x,y,z\,(\succ(x,y,z)\rightarrow \vert(y)) \\
\forall x,y,z\,(\succ(x,y,z)\rightarrow \exists w\,(\succ(x,z,w))) \\
\forall x\,(\vert(x)\rightarrow \exists y\,(\succ(x,0,y)))
\end{array}
\]

Now let $G=\tup{V,E}$ be a directed graph ($V$ is the set of vertices of $G$ and $E$ is the set of edges of $G$) and let $s,t\in V$. W.l.o.g.\ we assume that $G$ is represented through an adjacency list and that $0\not\in V$.
We define the set of facts $\D$ as follows: 
\begin{tabbing}
$\qquad\D=\ $\=$\{ \vert(a) \mid a\in V \wedge a\neq t \} \;\cup$ \\
\>$\{ \succ(a,0,0) \mid a\in V \wedge \textit{ a has no successors in } G \} \;\cup$ \\
\>$\{ \succ(a,0,b_1),\succ(a,b_1,b_2),\ldots,\succ(a,b_{h-1},b_h),\succ(a,b_h,0) \mid$ \\
\>$\qquad a\in V \wedge\tup{b_1,\ldots,b_h} \textit{ is the adjacency list of } a \}$
\end{tabbing}

Observe that $\Sigma$ is a set of \ffk dependencies for $\D$ (in particular, for every $x$ there is at most one fact in $\D$ of the form $\succ(x,0,\cdot)$, and for every $x,y$ there is at most one fact in $\D$ of the form $\succ(x,y,\cdot)$).
Finally, let $\D'=\{\vert(s)\}$.

It is possible to verify that there exists a path in $G$ from $s$ to $t$ iff $\D'$ is not weakly consistent with $\SD$. The key point is that $\Sigma$ is such that: $(i)$ all the incoming edges of $t$, i.e., all the facts of the form $\succ(\cdot,t,\cdot)$, must be deleted in all repairs due to the first dependency and the absence of $\vert(t)$ in $\D$; $(ii)$ due to the second dependency, the elimination of one edge $(a,b)$ represented by the fact $\succ(a,b,n)$ implies the elimination of all the outgoing edges of $a$, i.e.\ the elimination of all the facts of the form $\succ(a,\cdot,\cdot)$. This in turn implies, by the third dependency, the elimination of the fact $\vert(a)$. Consequently, for every vertex $a$, $\vert(a)$ belongs to the only repair of $\SD$ iff $t$ is not reachable from $a$ in $G$.
\end{proof}
\fi 

\iflong 
\begin{theorem}
\label{thm:wc-lb-ffk}
Weak consistency in the case of \ffk dependencies is PTIME-hard \wrt data complexity.
\end{theorem}
\begin{proof}
We prove the thesis by reduction from HORN 3-SAT (a well-known PTIME-complete problem). We define the following set of \ffk (actually, full) dependencies $\Sigma$:
\[
\begin{array}{l}
\forall x\,(C(0,0,x)\rightarrow A(x)) \\
\forall x,y\,(C(x,0,y)\wedge A(x)\rightarrow A(y)) \\
\forall x,y,z\,(C(x,y,z)\wedge A(x)\wedge A(y)\rightarrow A(z)) \\
\forall x\,(C_f(x,0,0)\wedge A(x)\rightarrow \bot) \\
\forall x,y\,(C_f(x,y,0)\wedge A(x)\wedge A(y)\rightarrow \bot) \\
\forall x,y,z\,(C_f(x,y,z)\wedge A(x)\wedge A(y)\wedge A(z)\rightarrow \bot)
\end{array}
\]
\
Then, given a Horn 3-CNF $\phi$, we define $\D'$ as the set containing the following facts:
\begin{itemize}
\item
$C(0,0,a)$ for each clause of the form $a$ in $\phi$;
\item
$C(a,0,b)$ for each clause of the form $\neg a \vee b$ in $\phi$;
\item
$C(a,b,c)$ for each clause of the form $\neg a \vee \neg b \vee c$ in $\phi$;
\item
$C_f(a,0,0)$ for each clause of the form $\neg a$ in $\phi$;
\item
$C_f(a,b,0)$ for each clause of the form $\neg a \vee \neg b$ in $\phi$;
\item
$C_f(a,b,c)$ for each clause of the form $\neg a \vee \neg b \vee \neg c$ in $\phi$.
\end{itemize}
Moreover, let $V$ be the set of propositional variables occurring in $\phi$. We define the set of facts $\D''=\{ A(a) \mid a\in V\}$. Finally, let $\D=\D'\cup\D''$.

It can be shown that the models of $\phi$ are in 1-to-1 correspondence with the repairs of $\SD$ containing $\D'$. Consequently, $\phi$ is satisfiable iff there exists a repair of $\SD$ containing $\D'$, which implies that $\phi$ is satisfiable iff $\D'$ is weakly consistent with $\SD$.
\end{proof}

\begin{theorem}
\label{thm:wc-lb-linear}
Weak consistency in the case of linear dependencies is PTIME-hard \wrt data complexity.
\end{theorem}
\begin{proof}
We prove the thesis by showing that the HORN SAT problem can be reduced to (non-)weak consistency.

Given a set of $n$ ground Horn rules (we represent rules without head with an extra variable $\false$ and assume w.l.o.g.\ that there is at least one rule without head whose id is $r_1$), we represent a rule $r_i$ of the form $a \leftarrow b_1,\ldots,b_k$ in the database $\D$ as follows:
$H(r_i,a,1,1)$ (if there are multiple rules $r_{i_1},\ldots,r_{i_h}$ having $a$ in the head, we write $H(r_{i_1},a,1,2), H(r_{i_1},a,2,3), \ldots, H(r_{i_h},a,h,1)$), $B(r_i,b_1), \ldots, B(r_i,b_k)$.

We define the following set of dependencies $\Sigma$:
\[
\begin{array}{l}
\forall x,y,z,w \,(H(x,y,z,w)\rightarrow \exists v \, B(x,v)) \\
\forall x,y,z,w \,(H(x,y,z,w)\rightarrow \exists v,t \, H(v,y,w,t)) \\
\forall x,y\,(B(x,y)\rightarrow \exists z,w,v \, H(z,y,w,v))
\end{array}
\]
where:
the first dependency implies that if for a rule there are no more body atoms, the head atom of that rule is deleted;
the second dependency implies that if a head atom is deleted, all the head atoms with that variable are deleted;
the third dependency implies that if there are no head atoms for a variable, all the body atoms for that variable are deleted.

Therefore, it can be verified that the Horn formula is unsatisfiable iff the fact $H(r_1,\false,1,2)$ belongs to the only repair of $\SD$, i.e.\ iff the set  $\{H(r_1,\false,1,2)\}$ is weakly consistent with $\SD$.
\end{proof}
\fi 

\begin{theorem}
\label{thm:wc-lb-acyclic}
Weak consistency in the case of acyclic dependencies is NP-hard \wrt data complexity.
\end{theorem}
\iflong 
\begin{proof}
The proof is obtained through a reduction from the 3-CNF problem. Let $\Sigma$ be the following set of acyclic dependencies:

\[
\begin{array}{l}
\forall x\,(R(x)\rightarrow \exists y\, V(x,y)) \\
\forall w,x_1,x_2,x_3,y_1,y_2,y_3\,
(C(w,x_1,y_1,x_2,y_2,x_3,y_3) \wedge \\
\qquad V(x_1,y_1) \wedge V(x_2,y_2) \wedge V(x_3,y_3) \rightarrow \bot)
\end{array}
\]

Now, let $\phi=\bigwedge_1^n(\ell_i^1\vee \ell_i^2\vee\ell_i^3)$ be a 3-CNF formula, $A$ be the set of propositional variables occurring in $\phi$, and $\D$ be the following set of facts:
\[
\begin{array}{l}
\{ C(i,\propvar(\ell_i^1),\ntv(\ell_i^1),\propvar(\ell_i^2),\ntv(\ell_i^2),\propvar(\ell_i^3),\ntv(\ell_i^3)) \mid 1\leq i\leq n \}\ \cup \\
\qquad \{ R(a), V(a,0), V(a,1) \mid a\in A \}
\end{array}
\]
where each $\propvar(\ell_i^j)$ is the propositional variable appearing in the literal $\ell_i^j$ and each $\ntv(\ell_i^j)$ is 1 if the literal $\ell_i^j$ is negated and 0 otherwise.
Finally, let $\D'$ be the following set of facts:
\[
\begin{array}{l}
\{ C(i,\propvar(\ell_i^1),\ntv(\ell_i^1),\propvar(\ell_i^2),\ntv(\ell_i^2),\propvar(\ell_i^3),\ntv(\ell_i^3)) \mid  1\leq i\leq n \}\ \cup \\
\qquad \{ R(a) \mid a\in A \}
\end{array}\]
(note that both $\D$ and $\D'$ are inconsistent with $\Sigma$).

It is now easy to verify that $\D'$ is weakly consistent with $\SD$ iff $\phi$ is satisfiable.
Let us first consider the case in which $\phi$ is satisfiable. Then there exists a guess of the variables of $\phi$ such that each clause of $\phi$ is satisfied.
Such guess can be represented with a set $\D''$ of facts with predicate $V$ corresponding to the images of the head of the first dependency.
Moreover, since all the clauses of $\phi$ must be satisfied, each of them must contain at least one variable having a truth value equal to 0 if the literal is negated and 1 otherwise.
Therefore, the set $\D'\cup \D''$ does not contain any image of the body of the second dependency. Thus $\D'\cup\D''$ is a repair of $\SD$ and, consequently, $\D'$ is weakly consistent with $\SD$.

Conversely, if $\phi$ is unsatisfiable, then notice that every repair $\D'\cup\D''$ containing $\D'$ should be such that $\D''$ contains a fact $V(a,\cdot)$ for every propositional variable $a$ (otherwise $\D'\cup\D''$ would be inconsistent with the first dependency). On the other hand, since $\phi$ is unsatisfiable, each possible set $\D''$ representing a guess of the truth values of the variables of $\phi$ is such that $\D'\cup\D''$ is inconsistent with the second dependency. Therefore, $\D'$ is not weakly consistent with $\SD$.
\end{proof}
\fi 

\iflong
\subsection{Repair checking}
\fi
\label{subsec:lb-rc}

\iflong
We now focus on the lower bounds of the problem of checking if a given database is a repair. 
We prove the repair checking
\fi
problem to be NL-hard if the considered dependencies are both linear and \ffk for the given database, and PTIME-hard in the case when such dependencies are just linear. For what concerns the \ffk case, we lead the lower bound back to \cite[Theorem~5]{AK09}, which shows a PTIME-hardness result in the case when the set of constraints consists of full TGDs and EGDs, both of which are \ffk for each possible database instance. Furthermore, the results provided in \cite{CM05} allow us to claim that, in the general case, repair checking is coNP-hard.

%

\begin{proposition}
\label{pro:rc-lb-general-and-linear}
Repair checking in the case of arbitrary dependencies is coNP-hard \wrt data complexity, while in the case of linear dependencies is PTIME-hard \wrt data complexity.
\end{proposition}
\iflong 
\begin{proof}
For the case of arbitrary dependencies, the thesis is implied by Theorem 4.4 of \cite{CM05}.
For linear dependencies, the thesis follows immediately from Theorem~\ref{thm:ic-lb-linear} (see Section~\ref{subsec:lb-ic}) and from Theorem 4.1 of \cite{CM05} (which allows for extending the lower bound of query entailment to the complement of the repair checking problem).
\end{proof}
\fi

\begin{theorem}
\label{thm:rc-lb-linear-ffk}
Repair checking in the case of linear \ffk dependencies is NL-hard \wrt data complexity.
\end{theorem}
\iflong 
\begin{proof}
We prove the thesis by showing a reduction from STCON (the reachability problem on directed graphs).
Let $\Sigma$ be the following set of linear dependencies:
\[
\begin{array}{l}
\forall x,y,z\,(\succ(x,y,z)\rightarrow \vert(y)) \\
\forall x,y,z\,(\succ(x,y,z)\rightarrow \exists w\,(\succ(x,z,w))) \\
\forall x\,(\vert(x)\rightarrow \exists y\,(\succ(x,0,y))) \\
\forall x\,(\initial(x)\rightarrow \vert(x)) \\
\forall x\,(\vert(x)\rightarrow \exists y\,\initial(y)) \\
\forall x,y,z\,(\succ(x,y,z)\rightarrow \exists w\,\initial(w)) 
\end{array}
\]

Now, let $G=\tup{V,E}$ be a directed graph ($V$ is the set of vertices of $G$ and $E$ is the set of edges of $G$) and let $s,t\in V$. W.l.o.g.\ we assume that $G$ is represented through an adjacency list and that $0\not\in V$.
We define the following set of facts $\D$: 
\[
\begin{array}{l}
\{ \initial(s) \} \cup
\{ \vert(a) \mid a\in V \wedge a\neq t \} \;\cup \\
\{ \succ(a,0,0) \mid a\in V \wedge \textit{ a has no successors in } G \} \;\cup \\
\{ \succ(a,0,b_1),\succ(a,b_1,b_2),\ldots,\succ(a,b_{h-1},b_h),\succ(a,b_h,0) \mid \\
\quad\quad a\in V \wedge\tup{b_1,\ldots,b_h} \textit{ is the adjacency list of } a \} 
\end{array}
\]
Observe that $\Sigma$ is a set of \ffk dependencies for $\D$ (in particular, for every $x$ there is at most one fact in $\D$ of the form $\succ(x,0,\cdot)$, and for every $x,y$ there is at most one fact in $\D$ of the form $\succ(x,y,\cdot)$).
Finally, let $\D'=\emptyset$.

It is possible to verify that there exists a path in $G$ from $s$ to $t$ iff $\D'$ is a repair of $\SD$. The key point is that $\Sigma$ is such that: $(i)$ all the incoming edges of $t$, i.e., all the facts of the form $\succ(\cdot,t,\cdot)$, must be deleted in all repairs due to the first dependency and the absence of $\vert(t)$ in $\D$; $(ii)$ due to the second dependency, the elimination of one edge $(a,b)$ represented by the fact $\succ(a,b,n)$ implies the elimination of all the outgoing edges of $a$, i.e.\ the elimination of all the facts of the form $\succ(a,\cdot,\cdot)$. This in turn implies, by the third dependency, the elimination of the fact $\vert(a)$.
Therefore, if $t$ is reachable from $s$ in $G$, then $\vert(s)$ cannot belong to the repair of $\SD$, which by the fourth dependency implies that $\initial(s)$ cannot belong to the repair of $\SD$, which by the fifth and sixth dependency implies that no other facts of $\D$ cannot belong to the repair of $\SD$, i.e.\ the repair of $\SD$ is the empty set.
\end{proof}
\fi

\iflong
\subsection{Instance checking and query entailment}
\fi
\label{subsec:lb-ic}

\iflong 
We conclude this section turning our attention to the lower bounds for instance checking and for query entailment under both \irs\ and \ars\ semantics. 
\fi 
First, the results of \cite{CM05} imply that \ars-entailment is coNP-hard in the case of acyclic dependencies and in the case of \ffk dependencies, and instance checking is $\pidue$-hard in the general case. 
We now show that instance checking under both \irs\ and \ars\ semantics
are NL-hard if the considered dependencies are both linear and \ffk for the given database, PTIME-hard if such dependencies are just linear, and
coNP-hard when the given dependencies are \ffk (actually, full) or acyclic.

%

\iflong 
\begin{proposition}[\cite{CM05}, Thm.~4.7]
\label{pro:ic-lb-general}
Instance checking in the case of arbitrary dependencies is $\pidue$-hard \wrt data complexity.
\end{proposition}


\begin{proposition}[\cite{CM05}, Thm.~3.3]
\label{pro:ar-lb-acyclic-ffk}
CQ entailment under \ars\ semantics in the case of acyclic \ffk dependencies is coNP-hard \wrt data complexity.
\end{proposition}
\fi 

\iflong
\begin{theorem}
\label{thm:ic-lb-acyclic}
Instance checking in the case of acyclic dependencies is coNP-hard w.r.t.\ data complexity.
\end{theorem}

\begin{proof}
The thesis can be proved through a reduction from the 3-CNF problem.
Let $\Sigma$ be the following set of acyclic dependencies:
\[
\begin{array}{l}
\forall x\,(V(x,1)\wedge V(x,0)\rightarrow \bot) \\
\forall x,y_1,y_2,y_3,z_1,z_2,z_3\,
(C_1(x,y_1,z_1,y_2,z_2,y_3,z_3)\wedge\\
\qquad V(y_1,z_1)\wedge V(y_2,z_2)\wedge V(y_3,z_3) \rightarrow \bot) \\
\ldots \\
\forall x,y_1,y_2,y_3,z_1,z_2,z_3\,
(C_7(x,y_1,z_1,y_2,z_2,y_3,z_3)\wedge\\
\qquad V(y_1,z_1)\wedge V(y_2,z_2)\wedge V(y_3,z_3) \rightarrow \bot) \\
\forall x,y_1,y_2,y_3,y_4,y_5,y_6\,
(C_1(x,y_1,y_2,y_3,y_4,y_5,y_6) \rightarrow \\
\qquad
\exists w_1,w_2,w_3,w_4,w_5,w_6\: C_2(x,w_1,w_2,w_3,w_4,w_5,w_6)) \\
\ldots \\
\forall x,y_1,y_2,y_3,y_4,y_5,y_6\,
(C_6(x,y_1,y_2,y_3,y_4,y_5,y_6) \rightarrow \\
\qquad
\exists w_1,w_2,w_3,w_4,w_5,w_6\: C_7(x,w_1,w_2,w_3,w_4,w_5,w_6)) \\
U \rightarrow \exists x,y_1,y_2,y_3,y_4,y_5,y_6\: C_1(x,y_1,y_2,y_3,y_4,y_5,y_6)
\end{array}
\]
Then, let $\D$ be the following database that contains:
$V(a,0),V(a,1)$ for each variable $a$ occurring in $\phi$;
7 facts $C_1, \ldots, C_7$ for each clause of $\phi$ (each such fact represents an evaluation of the three variables of the clause that make the clause true);
and the fact $U$.

E.g. if the $i$-th clause is $\neg a \vee \neg b \vee c$, then $\D$ contains the facts
\[
\begin{array}{l}
C_1(i,a,0,b,0,c,0),
C_2(i,a,0,b,0,c,1),
C_3(i,a,0,b,1,c,0),
C_4(i,a,0,b,1,c,1),\\
C_5(i,a,1,b,0,c,0),
C_6(i,a,1,b,0,c,1),
C_7(i,a,1,b,1,c,1)
\end{array}
\]

We prove that the fact $U$ belongs to all the repairs of $\SD$ iff $\phi$ is unsatisfiable.

In fact, if $\phi$ is unsatisfiable, then there is no repair of $\SD$ without any fact of $C_1$, consequently $U$ belongs to all the repairs of $\SD$.
Conversely, if $\phi$ is satisfiable, then there is a repair (where the extension of $V$ corresponds to the interpretation satisfying $\phi$) that for each clause does not contain at least one of the 7 facts $C_1,\ldots,C_7$ representing the clause. Due to the dependencies between $C_i$ and $C_{i+1}$, this implies that there is a repair that for each clause does not contain any fact $C_1$, and therefore (due to the last dependency) it does not contain the fact $U$.
\end{proof}

\begin{theorem}
\label{thm:ic-lb-linear}
Instance checking in the case of linear dependencies is PTIME-hard w.r.t.\ data complexity.
\end{theorem}
\begin{proof}
The proof immediately follows from the reduction shown in the proof of Theorem~\ref{thm:wc-lb-linear}, which already shows that, given $\Sigma$ and $\D$ defined as in that proof, the Horn formula is unsatisfiable iff the fact $H(r_1,\false,1,2)$ belongs to the only repair of $\SD$.
\end{proof}

\begin{theorem}
\label{thm:ic-lb-linear-ffk}
Instance checking in the case of linear \ffk dependencies is NL-hard w.r.t.\ data complexity.
\end{theorem}
\begin{proof}
The proof immediately follows from the reduction shown in the proof of Theorem~\ref{thm:rc-lb-linear-ffk}: it is immediate to see that, in that proof, there exists a path in $G$ from $s$ to $t$ iff the fact $\initial(s)$ does not belong to the only repair of $\SD$.
\end{proof}
\fi

We now turn our attention to \ffk dependencies.
The following property actually holds already for the smaller class of \emph{full} dependencies, and extends a previous result for query entailment under \ars\ semantics (\cite{CFK12} Theorem 5.5).

\begin{theorem}
\label{thm:ic-lb-ffk}
Instance checking in the case of full dependencies is coNP-hard \wrt data complexity.
\end{theorem}
\iflong 
\begin{proof}
The proof is by reduction from 3-CNF.
We define the following set $\Sigma$ of full dependencies:

\[
\begin{array}{l}
\forall x_1,x_2,x_3,v_1,v_2,v_3,y,z\,\\
\qquad 
(S(z)\wedge N(y,z)\wedge C(y,x_1,v_1,x_2,v_2,x_3,v_3)\wedge V(x_1,y_1)\rightarrow S(y)) \\
\forall x_1,x_2,x_3,v_1,v_2,v_3,y,z\,\\
\qquad 
(S(z)\wedge N(y,z)\wedge C(y,x_1,v_1,x_2,v_2,x_3,v_3)\wedge V(x_2,y_2)\rightarrow S(y)) \\
\forall x_1,x_2,x_3,v_1,v_2,v_3,y,z\,\\
\qquad 
(S(z)\wedge N(y,z)\wedge C(y,x_1,v_1,x_2,v_2,x_3,v_3)\wedge V(x_3,y_3)\rightarrow S(y)) \\
\forall x\,(V(x,0)\wedge V(x,1)\rightarrow u) \\
S(1) \rightarrow u
\end{array}
\]

Given a 3-CNF formula $\phi$, in the database $\D$, we represent every clause of $\phi$ with (at most) 7 facts corresponding to the interpretations of the variables that satisfy the clause: e.g.\ if clause $n$ is $a\vee \neg b \vee c$, we add the facts

\[
\begin{array}{l}
C(n,a,0,b,0,c,0),C(n,a,0,b,0,c,1),C(n,a,0,b,1,c,1),\\
C(n,a,1,b,0,c,0),C(n,a,1,b,0,c,1),C(n,a,1,b,1,c,0),\\
C(n,a,1,b,1,c,1)
\end{array}
\]

Moreover, the database contains the facts $V(p,0),V(p,1)$ for every propositional variable $p$, and the facts 

\[
N(a_1,a_2),\ldots,N(a_{m-1},a_m),N(a_m,a),
S(a_1),\ldots,S(a_m),S(a)
\]

We prove that $\phi$ is unsatisfiable iff the fact $S(a)$ belongs to all the repairs of $\SD$.

First, if $\phi$ is satisfiable, then let $P$ be the set of propositional variables occurring in $\phi$, let $I$ be an interpretation (subset of $P$) satisfying $\phi$, and let $\D'$ be the following subset of $\D$:

\[
\D' = \D \setminus (
    \{V(p,0) \mid p\in P\cap I \}\;\cup
    \{ V(p,1) \mid p\in P\setminus I \} \cup \{S(a)\} )
\]

It is immediate to verify that $\D'$ is consistent with $\Sigma$ and that $\D'\cup \{s(a)\}$ is not weakly consistent with $\SD$: in fact, the addition of $S(a)$ creates a sequence of instantiations of the bodies of the first three dependencies of $\Sigma$ that requires (to keep the consistency of the set) to add to $\D'\cup\{s(a)\}$ first the fact $s(a_m)$, then $s(a_{m-1})$, and so on until $s(a_1)$,but this makes this set inconsistent with $\Sigma$, due to the last dependency of $\Sigma$ and the absence of the fact $u$ in $\D$. Consequently, there exists a repair of $\SD$ that does not contain $s(a)$.

On the other hand, if $\phi$ is unsatisfiable, then every guess of the atoms of the $V$ predicate that satisfies the fourth dependency (each such guess corresponds to an interpretation of the propositional variables) is such that the sequence of instantiations of the bodies of the first three dependencies of $\Sigma$ mentioned above, which leads to the need of adding $S(a_1)$ to the set, is blocked by the absence of some fact for $V$. Consequently, the deletion of $S(a_1)$ does not imply the deletion of $S(a)$ and $S(a)$ belongs to all the repairs of $\SD$.
\end{proof}
\fi

%% file: 6-weak-consistency.tex
\newcommand{\algars}{\ars-CQEnt\xspace}
\newcommand{\algirs}{\irs-CQEnt\xspace}
\newcommand{\algacyclicffk}{Acyclic-\ffk-\irs-CQEnt\xspace}
\newcommand{\algweaklyconsistentacyclic}{Check-Weak-Consistency-Acyclic\xspace}
\newcommand{\apply}{\mathit{Apply}}
\newcommand{\fccons}{\phi_{FC}}
\newcommand{\phieq}{\phi_{eq}}
\newcommand{\notunifiable}{\mathit{NotUnif}}
\newcommand{\headsequalities}{\textit{HE}}
\newcommand{\aeq}{\A_{eq}}
\newcommand{\Tau}{\Sigma_T'}
\newcommand{\seq}{\textit{Seq}}

\newcommand{\inconsistent}{\textit{Incons}}
\newcommand{\inconsistentprimed}{\textit{IncAux}}
\newcommand{\weaklyinconsistent}{\textit{WeakIncons}}
\newcommand{\rewr}{\textit{Rewr}}
\newcommand{\applicable}{\textit{Applicable}}
\newcommand{\primed}{\textit{Primed}}
\newcommand{\queryrewrite}{\textit{QEnt}}
\newcommand{\atomsets}{\textit{AtomSets}}
\newcommand{\algcomputerepairlinear}{Compute-Repair-Linear\xspace}
\newcommand{\algwcffk}{Check-Weak-Consistency-\ffk\xspace}
\newcommand{\forwardclosure}{forward closure\xspace}
\newcommand{\belongstofc}{\textit{InFC}}
\newcommand{\weaklycons}{\textit{WCons}}
\newcommand{\weaklyconsacycliclinear}{\textit{WCons\textsuperscript{AL}}}
\newcommand{\headunify}{\textit{HeadUnify}}
\newcommand{\bodyunify}{\textit{BodyUnify}}

\section{Checking weak consistency}
\label{sec:wc}

In this section we establish the upper bounds of the problem of checking weak consistency. It is easy to see that, in the general case, the problem belongs to NP.
Anyway, the problem becomes tractable (indeed, it is in $\aczero$) if the acyclic property is accompanied by the \ffk or linear one. In order to prove this result, we provide two first-order rewriting techniques.
We also prove that the problem can be solved in polynomial time if the considered dependencies are linear or \ffk for the given database and, more specifically, it belongs to NL if both such properties are enjoyed.

%


First, we recall
\iflong the following property for the consistency problem.
\fi
\begin{proposition}
\label{pro:consistency-ub}
Deciding consistency of a database with a set of dependencies is $\aczero$ \wrt data complexity.
\end{proposition}
\iflong 
\begin{proof}
The proof immediately follows from the fact that a database $\D$ is consistent with $\Sigma$ iff $\D\models\bigwedge_{\tau\in\Sigma}\tau$, and from the fact that evaluating a domain-independent FO sentence over a database is in $\aczero$ \wrt data complexity. 
\end{proof}
\fi 

From the above property and from the definition of weak consistency, it immediately follows that:

\begin{proposition}
\label{pro:wc-ub-general}
In the general case (arbitrary dependencies), deciding weak consistency is NP \wrt data complexity.
\end{proposition}



We now turn our attention to \ffk dependencies. For this class of dependencies, we define the notion of \emph{\forwardclosure} of a subset of the database as follows. 
Let $\Sigma$ be a set of dependencies and let $\D$ be a database such that $\Sigma$ is \ffk for $\D$. We denote by $\fc(\D',\D,\Sigma)$ the minimal superset of $\D'$ such that,
if there exists a dependency $\tau\in\Sigma$ and an instantiation $\sigma$ of $\body(\tau)$ in $\D''$ and there exists an image $M$ of $\head(\sigma(\tau))$ in $\D$, 
then $M\subseteq\D''$.

Now, in the case of a \ffk dependency $\tau$, for each instantiation $\sigma$ of  $\body(\tau)$ there exists at most one image of $\head(\sigma(\tau))$ in $\D$. Consequently, the following property holds.

\begin{proposition}
\label{pro:wc-fc}
Let $\Sigma$ be a set of dependencies, let $\D$ be a database such that $\Sigma$ is \ffk for $\D$, and let $\D'\subseteq\D$. Then, $\D'$ is weakly consistent with $\SD$ iff $\fc(\D',\D,\Sigma)$ is consistent with $\Sigma$.
\end{proposition}

We are now ready to establish the complexity of weak consistency in the case of \ffk dependencies.




\begin{theorem}
\label{thm:wc-ub-ffk}
When $\Sigma$ is a set of \ffk dependencies for $\D$, deciding the weak consistency of a subset $\D'$ of $\D$ with $\SD$ is PTIME \wrt data complexity. 
\end{theorem}
\iflong 
\begin{proof}
The proof follows from Proposition~\ref{pro:wc-fc} and from the fact that $\fc(\D'\,\D,\Sigma)$ can be computed in polynomial time \wrt data complexity through an iterative algorithm that, for every instantiation of the body of a dependency in $\D'$, adds to $\D'$ the corresponding (unique) image of the head of the dependency in $\D$, until a fixpoint is reached (and a fixpoint is reached after at most $n$ additions of new images of dependency heads, if $n$ is the number of facts in $\D$).
\end{proof}
\fi  

We now consider acyclic \ffk dependencies.
First, we prove that, for such dependencies, it is possible to build an FO sentence that decides whether an atom belongs to $\fc(\D',\D,\Sigma)$. Then, we use such a sentence as a subformula of an FO sentence that decides whether $\D'$ is weakly consistent with $\SD$.

To this aim, in these formulas, besides the predicates $\{p_1,\ldots,p_m\}$ used in $\SD$, we make use of a second, auxiliary set of predicates $\{p_1^\aux,\ldots,p_m^\aux\}$, that we use to represent the subset $\D'$ of $\D$. 
Given a formula $\phi$, we denote by $\aux(\phi)$ the formula obtained from $\phi$ replacing every predicate atom $p(t)$ with $p^\aux(t)$, i.e.\ the atom obtained from $\alpha$ by replacing its predicate with the corresponding predicate in the auxiliary alphabet
(and given a set of atoms $\A$, we denote by $\aux(\A)$ the set $\{\aux(\alpha)\mid \alpha\in\A\}$). 


Given a predicate atom $\alpha$, a set of acyclic dependencies $\Sigma$, where $\tup{\tau_1,\ldots,\tau_h}$ is a topological order of $\Sigma$, and an integer $i$, we recursively define $\belongstofc(\alpha,\Sigma,i)$ as follows: $(i)$ if $i=0$, then $\belongstofc(\alpha,\Sigma,i)$ is the formula $\aux(\alpha)$; $(ii)$ if $\leq i\leq h$, then $\belongstofc(\alpha,\Sigma,i)$ is the formula
\[
\begin{array}{r@{}l}
\displaystyle
\aux(\alpha)\vee 
\displaystyle
\bigvee_{\tau_j\in\{\tau_1,\ldots,\tau_i\}}
\bigvee_{\footnotesize\begin{array}{c}q\in\cq(\head(\tau_j)) \wedge\\
\headunify(\alpha,q,\sigma)\end{array}} 
\exists x \Big(&\sigma(\body(\tau_j))\wedge\sigma(\cnj(q))\wedge \\&
\displaystyle
\bigwedge_{\footnotesize\beta\in\bodyp(\sigma(\tau_j))}\belongstofc(\beta,\Sigma,j-1)\Big)
\end{array}
\]
where $\headunify(\alpha,q,\sigma)$
is true iff 
the atom $\alpha$ unifies with some atom of $q$ with MGU $\sigma$, and $x$ is the sequence of all the variables of $\tau_j$ that have not been substituted by $\sigma$.\footnote{In this definition, as well as in all the subsequent definitions of first-order sentences of this section, we make the usual assumption that every dependency involved in the definition of the sentence uses ``fresh" variables, i.e.\ variable symbols not already occurring in the sentence.}

\iflong 
To prove the correctness of the above formula, we need the following property, whose proof is straightforward:

\begin{lemma}
\label{lem:belongstofc-aux}
Let $\Sigma$ be a set of acyclic dependencies, where $\tup{\tau_1,\ldots,\tau_h}$ is a topological order of $\Sigma$. For every fact of the form $p(t)$, $\alpha$ be an atom and let $\sigma$ be a substitution of the variables occurring in $\alpha$ with constants. Then, for every database $\D$ such that $\Sigma$ is \ffk for $\D$, and for every $\D'\subseteq\D$, $p(t)\in\fc(\D',\D,\Sigma)$ iff $p(t)\in\fc(\D',\D,\{\tau_1,\ldots,\tau_j\})$, where $j$ is the largest integer such that the predicate $p$ occurs in the head of $\tau_j$. 
\end{lemma}

We are now ready to prove the crucial property of the formula $\belongstofc(\alpha,\Sigma,i)$.
\fi

\begin{lemma}
\label{lem:belongstofc}
Let $\Sigma$ be a set of acyclic dependencies, where $\tup{\tau_1,\ldots,\tau_h}$ is a topological order of $\Sigma$, let $\alpha$ be an atom and let $\sigma$ be a substitution of the variables occurring in $\alpha$ with constants. Then, for every database $\D$ such that $\Sigma$ is \ffk for $\D$, for every $\D'\subseteq\D$, and for every $i$ such that $0\leq i\leq h$, $\sigma(\alpha)\in\fc(\D',\D,\{\tau_1,\ldots,\tau_i\})$ iff $\D\cup\aux(\D')\models\sigma(\belongstofc(\alpha,\Sigma,i))$.
\end{lemma}

\iflong 
\begin{proof}
First, in the case when $i=0$, $\belongstofc(\alpha,\Sigma,i)=\aux(\alpha)$, and of course $\fc(\D',\D,\emptyset)=\D'$, and since $\sigma(\alpha)\in\D'$ iff $\D\cup\aux(\D')\models\sigma(\aux(\alpha))$, the thesis follows.

Now suppose that the thesis holds for every $i$ such that $i<\ell$, and consider the case when $i=\ell$.
From the definition of forward closure it follows that $\sigma(\alpha)$ belongs to $\fc(\D',\D,\{\tau_1,\ldots,\tau_i\})$ iff one of these two conditions holds:
\begin{itemize}
\item[$(i)$] $\sigma(\alpha)\in\D'$;
\item[$(ii)$] there exists $\tau_j\in\{\tau_1,\ldots,\tau_i\}$, $q\in\cq(\head(\tau_j))$ and an instantiation $\sigma'$ of $\body(\tau_j)$ such that $\headunify(\sigma(\alpha),q,\sigma')$ is true and $\sigma(\alpha)$ belongs to the (unique) image of $\head(\sigma'(\tau_j))$ in $\D$ and $\bodyp(\sigma'(\tau_j))\subseteq\fc(\D',\D,\{\tau_1,\ldots,\tau_i\})$.
Moreover, Lemma~\ref{lem:belongstofc-aux} implies that $\bodyp(\sigma'(\tau_j))\subseteq\fc(\D',\D,\{\tau_1,\ldots,\tau_i\})$ iff $\bodyp(\sigma'(\tau_j))\subseteq\fc(\D',\D,\{\tau_1,\ldots,\tau_{j-1}\})$.
\end{itemize}

Now, as already mentioned, $\D\cup\aux(\D')\models\sigma(\aux(\alpha))$ iff the above condition $(i)$ holds (notice that $\sigma(\aux(\alpha))$ is the first disjunct of $\sigma(\belongstofc(\alpha,\Sigma,i))$).

Moreover, let $\psi$ be the second disjunct of $\sigma(\belongstofc(\alpha,\Sigma,i))$, i.e.:
\[
\begin{array}{r@{}l}
\displaystyle
\bigvee_{\tau_j\in\{\tau_1,\ldots,\tau_i\}}
\bigvee_{\footnotesize\begin{array}{c}q\in\cq(\tau_j) \wedge\\
\headunify(\alpha,q,\sigma)\end{array}} 
\exists x \Big(&\sigma(\body(\tau_j))\wedge\sigma(\cnj(q))\wedge \\&
\displaystyle
\bigwedge_{\footnotesize\beta\in\bodyp(\sigma(\tau_j))}\belongstofc(\beta,\Sigma,j-1)\Big)
\end{array}
\]
It is immediate to verify that, due to the inductive hypothesis, $\D\cup\aux(\D')\models\psi$ iff the above condition $(ii)$ holds.
Consequently, the thesis follows.
\end{proof}
\fi 

We now define the sentence $\weaklycons(\Sigma)$ as follows:
\[
\begin{array}{r@{}l}
\displaystyle
\bigwedge_{\tau\in\Sigma} \forall x\,\Big(&
\displaystyle
\Big(\bigwedge_{\footnotesize\begin{array}{l}\alpha\in\predatoms(\body(\tau))\\ \;\end{array}} \belongstofc(\alpha,\Sigma,1) \wedge\inequalities(\body(\tau))\Big) \ra \\&
\displaystyle
\Big(\bigvee_{q\in\cq(\head(\tau))} \exists y\,\big(\big(\bigwedge_{\beta\in\predatoms(q)} \belongstofc(\beta,\Sigma,1)\big)\wedge\inequalities(q)\big)\Big)\Big)
\end{array}
\]
where $x$ is the sequence of the universally quantified variables of $\tau$ and $y$ is the sequence of the existentially quantified variables of $q$.

\begin{theorem}
\label{thm:weaklycons}
Let $\Sigma$ be a set of acyclic dependencies, where $\tup{\tau_1,\ldots,\tau_h}$ is a topological order of $\Sigma$. For every database $\D$ such that $\Sigma$ is \ffk for $\D$, and for every $\D'\subseteq\D$, $\D'$ is weakly consistent with $\SD$ iff $\D\cup\aux(\D')\models\weaklycons(\Sigma)$. 
\end{theorem}
\iflong 
\begin{proof}
First, observe that 
$\D'$ is weakly consistent with $\SD$ iff 
$\fc(\D',\D,\Sigma)$ is consistent with $\Sigma$, i.e.\ for every dependency $\tau$ and every instantiation $\sigma$ of $\body(\tau)$, there exists $q\in\cq(\head(\tau))$ such that $q$ has an image in $\fc(\D',\D,\Sigma)$.
Then, the thesis follows immediately from the definition of $\weaklycons(\Sigma)$ and 
from Lemma~\ref{lem:belongstofc}.
\end{proof}
\fi 

\begin{corollary}\label{cor:wc-ub-acyclic-ffk}
Let $\Sigma$ be an acyclic \ffk set of dependencies for $\D$ and let $\D'\subseteq\D$. Checking weak consistency of $\D'$ with $\SD$ is $\aczero$ \wrt data complexity.
\end{corollary}

\begin{example}\label{ex:wc-acyclic-ffk}
Let
$\D=\{ P(a,b), T(a,c), R(a,d), P(e,f), T(e,g),\allowbreak R(e,e)\}$ and
$\constr=\{
    \forall v\,(R(v,v) \rightarrow \bot),
    \forall x,y,z\, (P(x,y) \wedge T(x,z) \rightarrow (\exists w\, R(x,w) \wedge w \ne z))
\}$.
Note that $\constr$ is an acyclic set of \ffk dependencies for $\D$ and, in particular, the second dependency is not linear.
The sentence $\phi=\weaklycons(\Sigma)$ is equal to:
\[
\begin{array}{l}
\forall x,y,z\,\big((P^\aux(x,y) \wedge T^\aux(x,z)) \ra (\exists w R^\aux(x,w) \wedge w \ne z)\big) \wedge\\
\forall v\,\big(\big(R^\aux(v,v) \vee (\exists y,z\,(P(v,y) \wedge T(v,z) \wedge R(v,v) \wedge v \ne z \wedge P^\aux(v,y) \wedge T^\aux(v,z)))\big) \ra \bot\big)
\end{array}
\]

Let us now consider two subsets of $\D$, namely $\D'=\{P(a,b), T(a,c)\}$ and $\D''=\{P(e,f), T(e,g)\}$. One can verify that $\D\cup\aux(D')\models\phi$ and $\D\cup\aux(D'')\not\models\phi$. Indeed, we have that $\D'$ is weakly consistent with $\SD$, while $\D''$ is not.
\end{example}

For linear dependencies, we first present the algorithm \algcomputerepairlinear (Algorithm~\ref{alg:compute-repair-linear}) that, given a set of linear dependencies $\Sigma$ and a database $\D$, computes the unique repair of $\SD$.


\begin{algorithm}
\caption{\algcomputerepairlinear}
\label{alg:compute-repair-linear}
\begin{algorithmic}[1]
\REQUIRE A database $\D$, a set of dependencies $\constr$;
\ENSURE A database $\D'$;
\STATE let $\D'=\D$; \\
\STATE \textbf{repeat} \\
\STATE \quad \textbf{for each} dependency $\tau\in\Sigma$ \textbf{do} \\
\STATE \qquad \textbf{for each} instantiation $\sigma$ of $\body(\tau)$ in $\D'$ \\
\STATE \qquad such that $\head(\sigma(\tau))$ has no image in $\D'$ \textbf{do} \\
\STATE \qquad\quad $\D'=\D'\setminus\bodyp(\sigma(\tau))$; \\
\STATE \textbf{until} a fixpoint for $\D'$ is reached; \\
\STATE \textbf{return} $\D'$ \\
\end{algorithmic}
\end{algorithm}


It is immediate to verify that the algorithm is correct and runs in PTIME.
It is also immediate to verify that a subset $\D'$ of $\D$ is weakly consistent with $\SD$ iff $\D'$ is a subset of the unique repair of $\SD$.
Consequently, we derive the following upper bound.

\begin{theorem}
\label{thm:wc-ub-linear}
Let $\Sigma$ be a set of linear dependencies, let $\D$ be a database and let $\D'\subseteq\D$. Deciding the weak consistency of $\D'$ with $\SD$ is PTIME \wrt data complexity. 
\end{theorem}

We now analyze the case of linear \ffk dependencies, and start by showing the following property (whose proof is straightforward) that holds for linear (not necessarily \ffk) dependencies.

\begin{lemma}
\label{lem:wc-linear}
Let $\Sigma$ be a set of linear dependencies, let $\D$ be a database, and let $\D'\subseteq\D$. Then, $\D'$ is weakly consistent with $\SD$ iff, for every $\alpha\in\D'$, $\{\alpha\}$ is weakly consistent with $\SD$.
\end{lemma}

We are now ready to prove the NL upper bound for the case of linear \ffk dependencies.

\begin{theorem}
\label{thm:wc-ub-linear-ffk}
When $\Sigma$ is a set of linear dependencies and $\D$ is a database such that $\Sigma$ is \ffk for $\D$, deciding the weak consistency of a subset $\D'$ of $\D$ with $\SD$ is NL \wrt data complexity. 
\end{theorem}
\iflong 
\begin{proof}
Using Lemma~\ref{lem:wc-linear}, we prove the thesis by reducing the problem of deciding the weak consistency of $\{\alpha\}$ with $SD$ to the STCON problem, through the following logspace reduction. Let $G$ be the directed graph such that: $(i)$ there is a vertex $\beta$ for every fact $\beta$ in $\D$; $(ii)$ there is an edge $(\beta,\gamma)$ in $G$ iff there exists a dependency $\tau\in\Sigma$ such that 
there exists an instantiation $\sigma$ of $\body(\tau)$ such that $\bodyp(\sigma(\tau))=\beta$ and $\gamma$ is an image of $\head(\sigma(\tau))$ in $\D$; $(iii)$ there is a further vertex $\bot$ in $G$ and there exists an edge $(\beta,\bot)$ in $G$ iff there exists a dependency $\tau\in\Sigma$ such that 
there exists an instantiation $\sigma$ of $\body(\tau)$ such that $\bodyp(\sigma(\tau))=\beta$ and there exists no image of $\head(\sigma(\tau))$ in $\D$.
It is now easy to verify that $\{\alpha\}$ is not weakly consistent with $\SD$ iff the vertex $\bot$ is reachable from the vertex $\alpha$ in $G$. Then, the thesis follows by Lemma~\ref{lem:wc-linear}.
\end{proof}
\fi 

Finally, we analyze the case of acyclic linear dependencies, and show that the weak consistency problem is in $\aczero$ for such a class of dependencies.
In a way similar to the case of acyclic \ffk dependencies, we define an FO sentence that decides weak consistency.

Given a set of acyclic linear dependencies $\Sigma$, where $\tup{\tau_1,\ldots,\tau_h}$ is a topological order of $\Sigma$, and an atom $\alpha$, we define the formula $\weaklyconsacycliclinear(\alpha,\Sigma)$ as follows: 
$(i)$ if $\Sigma=\emptyset$, then $\weaklyconsacycliclinear(\alpha,\Sigma)=\true$;
$(ii)$ if $\Sigma\neq\emptyset$, then $\weaklyconsacycliclinear(\alpha,\Sigma)$ is the formula
\[
\begin{array}{r@{}l}
\displaystyle
\bigwedge_{\footnotesize\begin{array}{c}\tau_i\in\Sigma \wedge\\\bodyunify(\alpha,\tau_i,\sigma)\end{array}} 
\Big(
\displaystyle
\bigvee_{q\in\cq(\head(\tau_i))} \exists x\,\big( \sigma(\cnj(q))\wedge 
\displaystyle
\bigwedge_{\beta\in\predatoms(q)}\weaklyconsacycliclinear(\sigma(\beta),\{\tau_{i+1},\ldots,\tau_{h}\}) \big) \Big)
\end{array}
\]
where $\bodyunify(\alpha,\tau_i,\sigma)$ is true iff the atom $\alpha$ unifies with $\body(\tau_i)$ with MGU $\sigma$, and $x$ is the sequence of all the variables of $\tau$ that have not been substituted by $\sigma$.%
\iflong\footnote{Notice that through the predicate $\bodyunify$ we are formalizing a form of forward chaining here, while the $\headunify$ predicate used in the case of acyclic \ffk dependencies formalizes a form of backward chaining.}\fi


\iflong 
\begin{lemma}
\label{lem:weaklyconsacycliclinear}
Let $\Sigma$ be a set of acyclic linear dependencies, 
let $\alpha$ be an atom and let $\sigma$ be a substitution of the variables occurring in $\alpha$ with constants. Then, for every database $\D$, 
$\sigma(\alpha)$ is weakly consistent with $\SD$ iff $\sigma(\alpha)\in\D$ and $\D\models\sigma(\weaklyconsacycliclinear(\alpha,\Sigma))$.
\end{lemma}
\fi 

Then, we define the formula $\weaklyconsacycliclinear(\Sigma)$ as follows:
\[
\begin{array}{l}
\displaystyle
\bigwedge_{p\in\pred(\Sigma)}\forall x\, (p^\aux(x)\rightarrow \weaklyconsacycliclinear(p(x),\Sigma))
\end{array}
\]

\begin{theorem}
\label{thm:wc-ub-acyclic-linear}
Let $\Sigma$ be a set of acyclic linear dependencies, let $\D$ be a database and let $\D'\subseteq\D$. Deciding the weak consistency of $\D'$ with $\SD$ is $\aczero$ \wrt data complexity. 
\end{theorem}
\iflong 
\begin{proof}
From Lemma~\ref{lem:weaklyconsacycliclinear}, it immediately follows that $\D'$ is weakly consistent with $\SD$ iff $\D\cup\aux(\D')\models\weaklyconsacycliclinear(\Sigma)$. Now, since $\weaklyconsacycliclinear(\Sigma)$ is an FO sentence, the thesis follows.
\end{proof}
\fi 

\begin{example}\label{ex:wc-acyclic-linear}
Let $\D=\{ P(a,b), T(b,c), T(a,d), T(a,e), R(a,d,b) \}$ and
$\constr=\{
    \forall x,y\, (P(x,y) \ra \exists z\, T(y,z) \wedge y \ne z),
    \forall x,y\, (T(x,y) \ra \exists v,w\, R(x,v,w))
\}$.
Note that $\Sigma$ is a set of acyclic linear dependencies and, in particular, the first one is not \ffk for $\D$.
The sentence $\phi=\weaklyconsacycliclinear(\Sigma)$ is equal to:
\[
\begin{array}{l}
\big(\forall x,y\,(P^\aux(x,y) \ra (\exists z\,(T(y,z)\wedge y \ne z \wedge \exists v,w\,R(y,v,w)))) \wedge\\
\quad \forall x,y\,(T^\aux(x,y) \ra \exists v,w\,R(x,v,w)) \wedge\\
\quad \forall x,y,z\,(R^\aux(x,y,z) \ra \top)\big)
\end{array}
\]
Let us now consider two subsets of $\D$, namely $\D'=\{T(a,e)\}$ and $\D''=\{P(a,b), T(b,c)\}$. One can verify that $\D\cup\aux(D')\models\phi$ and $\D\cup\aux(D'')\not\models\phi$. Indeed, we have that $\D'$ is weakly consistent with $\SD$, while $\D''$ is not.
\end{example}


%% file: 7-repair-checking.tex
\section{Repair checking}
\label{sec:rc}

\newcommand{\checkrepairacyclic}{\textit{CheckRepair}}

In this section we establish the upper bounds of the repair checking problem.
As shown in \cite[Proposition 4]{AK09}, in the general case, the problem is coNP.
We provide a first-order rewriting technique for proving the membership in $\aczero$ of the acyclic case.
We also show that the problem can be solved in PTIME if the dependencies are linear or \ffk for the given database, extending the same complexity result proved in \cite{SC10,CFK12} for less expressive dependencies. In particular, the problem is NL if both such properties are enjoyed.

%




First, in order to exploit the previous upper bounds for the weak consistency problem, we start by noticing the following property (whose proof is straightforward): given a set of dependencies $\Sigma$, a database $\D$ and a subset $\D'$ of $\D$, $\D'$ is a repair of $\SD$ iff $\D'$ is consistent with $\Sigma$ and there exists no $\alpha\in\D\setminus\D'$ such that $\D'\cup\{\alpha\}$ is weakly consistent with $\Sigma$.




Consequently: $(i)$ if deciding weak consistency is NP \wrt data complexity, repair checking is coNP \wrt data complexity; $(ii)$ if deciding weak consistency is PTIME \wrt data complexity, repair checking is PTIME \wrt data complexity.

Thus, Theorem~\ref{thm:wc-ub-linear} and Theorem~\ref{thm:wc-ub-ffk} imply the following property.

\begin{theorem}
\label{thm:rc-ub-linear-and-ffk}
Let $\Sigma$ be a set of dependencies, let $\D$ be a database and let $\D'\subseteq\D$. If $\Sigma$ is either a set of linear dependencies or an \ffk set of dependencies for $\D$, then checking whether $\D'$ is a repair of $\SD$ is PTIME \wrt data complexity.
\end{theorem}


We now consider the case of acyclic dependencies, and start by showing the following property.

\begin{theorem}
\label{thm:acyclic-layers}
Let $\Sigma$ be a set of acyclic dependencies, let $\D$ be a database, and let $\D'\subseteq\D$. $\D'$ is a repair of $\SD$ iff $\D'$ is consistent with $\Sigma$ and, for every $\alpha\in\D\setminus\D'$, $\D'\cup\{\alpha\}$ is inconsistent with $\Sigma$.
\end{theorem}
\iflong 
\begin{proof}
We prove by induction that, if $\D'$ is not a repair of $\SD$, then there exists $\alpha\in\D'\setminus\D$ such that $\D'\cup\{\alpha\}$ is consistent with $\Sigma$ (the other direction of the proof is straightforward). Let us stratify the set of predicates according to the dependencies (layer 1 = predicates not occurring in the body of any dependency; layer i = predicates occurring only in the body of dependencies such that only predicates of layers lower than i occur in the head). Let $\D_i,\D_i'$ denote respectively the projection of $\D$ and $\D'$ on the predicates of layer $i$. Now:

Base case: if $\D_1\neq\D_1'$, then there exists a fact $\alpha\in\D_1\setminus\D'$ such that $\D'\cup\{\alpha\}$ is consistent with $\Sigma$ (since $\alpha$ cannot generate new instantiations of bodies of dependencies because its predicate does not appear in any body of a dependency);

Inductive case: if $\D_j=\D_j'$ for each $i$ such that $\leq j\leq i$, then for every dependency $\tau$ of the $i$-th layer, every instantiation $\sigma$ of $\body(\tau)$ in $\D$ such that $\D\models\sigma(\head(\tau))$ is also such that $\D'\models\sigma(\head(\tau))$, consequently $\D'\cup\sigma(\body(\tau))$ is consistent with $\Sigma$, hence $\sigma(\body(\tau))\subseteq\D'$ (otherwise $\D'$ would not be a repair of $\SD$). Therefore, if $\alpha \in \D_{i+1}\setminus\D_{i+1}'$, then $\D'\cup\{\alpha\}$ is inconsistent with $\Sigma$.
\end{proof}
\fi 


Given a set $\Sigma$ of acyclic dependencies, we now define an FO sentence $\checkrepairacyclic(\Sigma)$ that decides repair checking. In a way similar to the cases of weak consistency for acyclic \ffk and acyclic linear dependencies, we use an auxiliary predicate $p^\aux$ for every predicate occurring in $\SD$.

First, we define the formula $\inconsistentprimed(\Sigma)$ as follows:
\[
\inconsistentprimed(\Sigma)=\bigvee_{\tau\in\Sigma} \exists x\, \big(\aux(\body(\tau))\wedge\neg(\\head(\tau)\big)
\]
where $x$ is the sequence of all the variables occurring in $\body(\tau)$.

Then, we define the auxiliary formula $\inconsistent(\Sigma,p(x))$, where $p(x)$ is an atom: it is obtained from the formula $\inconsistentprimed(\Sigma)$ by replacing every atom $p^\aux(t)$ 
 with the subformula $(p^\aux(t)) \vee (p(x) \wedge t=x))$ (or, equivalently, $p^\aux(t) \vee t=x)$ since in $\inconsistentprimed(\Sigma)$ atoms with auxiliary predicates only appear in the left-sides of implications).

We now define the formula $\checkrepairacyclic(\Sigma)$ as follows:
\[
\neg\inconsistentprimed(\Sigma) \wedge \bigwedge_{p\in\pred(\Sigma)}\forall x\,\big(p(x) \wedge \neg p^\aux(x) \rightarrow \inconsistent(\Sigma,p(x)\big)
\]
where $x=x_1,\ldots,x_m$ if $m$ is the arity of $p$.
From these definitions, the following property can be easily verified.

\begin{lemma}
Let $\Sigma$ be a set of acyclic dependencies, let $\D$ be a database and let $\D'\subseteq\D$. The sentence $\D\cup\aux(\D')\models\checkrepairacyclic(\Sigma)$ iff $\D'$ is a repair of $\SD$.
\end{lemma}

The above lemma immediately implies the following property.

\begin{theorem}
\label{thm:rc-ub-acyclic}
Let $\Sigma$ be a set of acyclic dependencies, let $\D$ be a database and let $\D'\subseteq\D$. Checking whether $\D'$ is a repair of $\SD$ is $\aczero$ \wrt data complexity.
\end{theorem}

\begin{example}
Let $\D=\{ P(a,a), T(a), R(a,b), R(a,c) \}$ and
$\constr=\{
    \forall x\, (P(x,x) \wedge T(x) \ra \exists y\, R(x,y))
\}$.
Note that $\Sigma$ is a set of acyclic dependencies and it is neither linear nor \ffk for $\D$.
With such an input, the sentence $\checkrepairacyclic(\Sigma)$ is equal to:
\[
\begin{array}{l}
\neg\big(\exists x,y\,(P^\aux(x,x) \wedge T^\aux(x) \wedge \neg R(x,y))\big) \wedge\\
\forall x',y'\,(P(x',y') \wedge \neg P^\aux(x',y') \ra \exists x,y\, ((P^\aux(x,x) \vee
\big(x=x' \wedge x= y')) \wedge \\
\qquad T^\aux \wedge \neg R(x,y))\big) \wedge\\
\forall x'\,(T(x) \wedge \neg T^\aux(x') \ra \exists x,y\, \big(P^\aux(x,x) \wedge (T^\aux \vee
x=x') \wedge \neg R(x,y))\big) \wedge\\
\forall x',y'\,\big(R(x',y') \wedge \neg R^\aux(x',y') \ra \exists x,y\, (P^\aux(x,x) \wedge
T^\aux \wedge \neg R(x,y))\big)
\end{array}
\]
One can verify that $\reps(\D) = \{ \{P(a,a), R(a,b), R(a,c)\}, \{T(a), R(a,b), \allowbreak R(a,c)\} \}$ and that these are the only two possible subsets of $\D$ such that $\D\cup\aux(\D')\models\checkrepairacyclic(\Sigma)$.
\end{example}







Finally, we consider the case of linear \ffk dependencies.

\begin{theorem}
\label{thm:rc-ub-linear-ffk}
Let $\Sigma$ be a set of linear dependencies, let $\D$ be a database such that $\Sigma$ is \ffk for $\D$ and let $\D'\subseteq\D$. Checking whether $\D'$ is a repair of $\SD$ is NL \wrt data complexity.
\end{theorem}
\iflong 
\begin{proof}
Given the hypothesis, it is immediate to verify that $\D'$ is a repair of $\SD$ iff: $(i)$ $\D'$ is consistent with $\Sigma$; $(ii)$ for every $\alpha\in\D\setminus\D'$, $\{\alpha\}$ is not weakly consistent with $\SD$. Since condition $(i)$ can be checked in $\aczero$ \wrt data complexity (by Proposition~\ref{pro:consistency-ub}) and condition $(ii)$ can be checked in NL \wrt data complexity (by Theorem~\ref{thm:wc-lb-linear-ffk}), the thesis follows.
\end{proof}
\fi 




%% file: 8-query-entailment.tex
\section{Query entailment}
\label{sec:qe}

\newcommand{\deltadue}{\Delta^p_2}
\newcommand{\queryentails}{\textit{QEnt\textsuperscript{AL}}}

In this section we establish the upper bounds of the entailment of safe BUCQs (which, throughout this section, we refer to simply as BUCQ) 
under both \ars\ and \irs\ semantics.

For arbitrary dependencies, we prove \ars-entailment to belong to $\pidue$, confirming the same upper bound proved in \cite{CFK12} for the case of BCQs without inequalities, and we obtain the same result for \irs-entailment.
Both query entailment problems become coNP if the given dependencies are acyclic or \ffk for the given database. We point out that the coNP-membership of \ars-entailment in the \ffk case extends the upper bound proved in \cite{CFK12} for GAV (i.e., full single-head) TGDs.
Moreover, it is interesting to see that, if the dependencies are both acyclic and \ffk for the given database, while for \ars-entailment the coNP upper bound can not be improved, \irs-entailment becomes instead tractable (indeed, it is in $\aczero$).
Finally, for the case of linear dependencies, we prove the two problems to be PTIME (same upper bound shown in \cite{CFK12} for the case of BCQs without inequalities) and, in particular, they are in $\aczero$ if also the acyclicity property is enjoyed and in NL if the considered dependencies are also \ffk for the given database.

\iflong
\subsection{Arbitrary dependencies and \ffk dependencies}
\fi
First, we present the algorithm \algars\ (Algorithm~\ref{alg:ars}) that, given a database $\D$, a set of arbitrary dependencies $\constr$, and a BUCQ $Q$, checks whether $Q$ is \ars-entailed by $\SD$.

\begin{algorithm}
\caption{\algars}
\label{alg:ars}
\begin{algorithmic}[1]
\REQUIRE A database $\D$, a set of dependencies $\constr$, a BUCQ $Q$;
\ENSURE A Boolean value;
\STATE \textbf{if} there exists $\D'\subseteq\D$ such that 
\STATE \quad $\D'$ is consistent with $\Sigma$ and $\D'\not\models Q$ and 
\STATE \quad for every $\alpha\in\D\setminus\D'$ 
\STATE \qquad $\D'\cup\{\alpha\}$ is not weakly consistent with $\SD$ 
\STATE \textbf{then return} false 
\STATE \textbf{else return} true
\end{algorithmic}
\end{algorithm}

\begin{proposition}
\label{pro:algorithm-ars-correct}
The algorithm \algars$(\Sigma,\D,Q)$ returns true iff the BUCQ $Q$ is \ars-entailed by $\SD$.
\end{proposition}
\iflong 
\begin{proof}
The proof immediately follows from the fact that a set $\D'$ satisfying the three conditions of the algorithm exists if and only if there exists a repair of $\SD$ that does not entail $Q$.
\end{proof}
\fi 


We are now ready to prove the following upper bounds for the query entailment problem under \ars\ semantics.

\begin{theorem}
\label{thm:ar-ub-general-and-ffk}
Entailment of BUCQs under \ars\ semantics is:
\iflong
\begin{itemize}
    \item[(a)] $\pidue$ \wrt data complexity in the general case of arbitrary dependencies;
    \item[(b)] coNP \wrt data complexity in the case of \ffk dependencies.
\end{itemize}
\fi
\end{theorem}
\iflong
\begin{proof}
  First, observe that consistency of a database \wrt a set of dependencies can be decided in $\aczero$ \wrt data complexity, as well as the evaluation of a BUCQ over a database. Then, the proof immediately follows from Algorithm~\ref{alg:ars}, Proposition~\ref{pro:algorithm-ars-correct}, Proposition~\ref{pro:wc-ub-general},
  and Theorem~\ref{thm:wc-ub-ffk}.
\end{proof}
\fi 


Then, we present the algorithm \algirs\ (Algorithm~\ref{alg:irs}) that, given a database $\D$, a set of arbitrary dependencies $\constr$, a BUCQ $Q$, checks whether $Q$ is \irs-entailed by $\SD$.

\begin{algorithm}
\caption{\algirs}
\label{alg:irs}
\begin{algorithmic}[1]
\REQUIRE A set of dependencies $\Sigma$, a database $\D$, a BUCQ $Q$;
\ENSURE A Boolean value;
\STATE \textbf{if} there exists an image $M$ of $Q$ in $\D$ such that
\STATE \quad there exists no $\D'\subseteq\D$ such that
\STATE \qquad $\D'$ is weakly consistent with $\Sigma$ in $\D$ and
\STATE \qquad $\D'\cup M$ is not weakly consistent with $\Sigma$ in $\D$
\STATE \textbf{then return} true
\STATE \textbf{else return} false
\end{algorithmic}
\end{algorithm}

The correctness of this algorithm follows easily from the definition of the \irs\ semantics.

\begin{proposition}
\label{pro:algorithm-irs-correct}
The algorithm \algirs$(\Sigma,\D,q)$ returns true iff the BUCQ $Q$ is \irs-entailed by $\SD$.
\end{proposition}
\iflong 
\begin{proof}
It is immediate to see that, if for an image $M$ there exists no subset $\D'$ of $\D$ satisfying the conditions of the algorithm, then there exists no repair of $\SD$ that does not contain $M$, therefore $Q$ is \irs-entailed by $\SD$. Conversely, if, for every image $M$ of the query, such a subset $\D'$ exists, then $\D'$ belongs to at least a repair $\R$ of $\SD$ which does not contain $M$, and therefore $M$ is not contained in the intersection of all the repairs of $\SD$, thus no image of $Q$ is contained in the intersection of all the repairs, which implies that $Q$ is not \irs-entailed by $\SD$.
\end{proof}
\fi


The above property implies the following upper bounds.

\begin{theorem}
\label{thm:ir-ub-general-and-ffk}
Entailment of BUCQs under \irs\ semantics is:
\iflong
\begin{itemize}
    \item[(a)] $\pidue$ \wrt data complexity in the case of arbitrary dependencies;
    \item[(b)] coNP \wrt data complexity in the case of \ffk dependencies.
\end{itemize}
\fi
\end{theorem}
\iflong
\begin{proof}
  First, observe that consistency of a database \wrt a set of dependencies can be decided in $\aczero$ \wrt data complexity, as well as the evaluation of a BUCQ over a database. Then, the proof immediately follows from Algorithm~\ref{alg:irs}, Proposition~\ref{pro:algorithm-irs-correct}, Proposition~\ref{pro:wc-ub-general},
  and Theorem~\ref{thm:wc-ub-ffk}.
\end{proof}
\fi

\iflong
\subsection{Acyclic dependencies}
\fi
We now prove the following upper bound for the case of acyclic dependencies.
\begin{theorem}
\label{thm:ar-ub-acyclic}
Let $\Sigma$ be a set of acyclic dependencies, let $\D$ be a database and let $Q$ be a BUCQ. Deciding whether $Q$ is \ars-entailed by $\SD$ is coNP \wrt data complexity. 
\end{theorem}
\iflong 
\begin{proof}
From the definition of \ars-entailment, it follows that a BUCQ $Q$ is not \ars-entailed by $\SD$ iff there exists a subset $\D'$ f $\D$ such that $(i)$ $\D'$ is a repair of $\SD$ and $(ii)$ $\D'\not\models Q$. Therefore, by Theorem~\ref{thm:rc-ub-acyclic} the thesis follows.
\end{proof}
\fi 

\begin{theorem}
\label{thm:ir-ub-acyclic}
Let $\Sigma$ be a set of acyclic dependencies, let $\D$ be a database and let $Q$ be a BUCQ. Deciding whether $q$ is \irs-entailed by $\SD$ is coNP \wrt data complexity.
\end{theorem}
\iflong 
\begin{proof}
First, observe that $Q$ is \irs-entailed by $\SD$ iff there exists an image of $Q$ in $\D$ such that $M\subseteq\intrep(\D)$, i.e.\ $M$ is contained in every repair of $\SD$.
Therefore, if $M_1,\ldots M_m$ are the images of $Q$ in $\D$, $Q$ is \irs-entailed by $\SD$ iff there exist $m$ repairs of $\SD$ $\R_1,\ldots,\R_m$ such that $M_i\not\subseteq\R_i$ for every $i$ such that $1\leq i\leq m$. Since the above number $m$ is bounded by $n^k$, where $n$ is the size of $\D$ and $k$ is the number of predicate atoms of $Q$, and by Theorem~\ref{thm:rc-ub-acyclic} repair checking in the case of acyclic dependencies is in $\aczero$ \wrt data complexity, the thesis follows.
\end{proof}
\fi

\iflong
\subsection{Linear dependencies}
\fi
For linear dependencies, we show the following property.

\begin{theorem}
\label{thm:qe-ub-linear}
Let $\Sigma$ be a set of linear dependencies, let $\D$ be a database and let $Q$ be a BUCQ. Deciding whether $Q$ is \ars-entailed by $\SD$ is PTIME \wrt data complexity. 
\end{theorem}
\iflong 
\begin{proof}
The proof follows immediately from Algorithm~\ref{alg:compute-repair-linear}: once computed (in PTIME) the only repair $\D'$ of $\SD$, the query is then evaluated over $\D'$ (which can be done in $\aczero$).
\end{proof}
\fi 



We recall that, in the case of linear dependencies, \ars-entailment and \irs-entailment of queries coincide.

\iflong
\subsection{Linear \ffk dependencies}
\fi
In the case of linear \ffk dependencies, we prove the following result.

\begin{theorem}
\label{thm:qe-ub-linear-ffk}
Let $\Sigma$ be a set of linear dependencies, let $\D$ be a database such that $\Sigma$ is a set of \ffk dependencies for $\D$, and let $Q$ be a BUCQ. Deciding whether $Q$ is \ars-entailed by $\SD$ is NL \wrt data complexity.
\end{theorem}
\iflong 
\begin{proof}
Given the hypothesis, it is immediate to verify that $Q$ is \ars-entailed by $\SD$ iff there exists an image $M$ of $Q$ in $\D$ such that, for every fact $\alpha\in M$, $\{\alpha\}$ is weakly consistent with $\SD$. Since there are at most $n^k$ images of $Q$ in $\D$, where $n$ is the size of $\D$ and $k$ is the number of predicate atoms of $Q$, and since by Theorem~\ref{thm:wc-lb-linear-ffk} weak consistency can be checked in NL \wrt data complexity, the thesis follows.
\end{proof}
\fi 

\iflong
\subsection{Acyclic \ffk dependencies}
\fi
\iflong We now consider the case of acyclic \ffk dependencies. \fi

First, we present the algorithm \algacyclicffk (Algorithm~\ref{alg:acyclic-ffk-irs}) to decide query entailment under \irs\ semantics for such a class of dependencies.

\begin{algorithm}
\caption{\algacyclicffk}
\label{alg:acyclic-ffk-irs}
\begin{algorithmic}[1]
\REQUIRE A set of acyclic dependencies $\Sigma$, a database $\D$ such that $\Sigma$ is \ffk for $\D$, a BUCQ $Q$;
\ENSURE A Boolean value;
\STATE let $k$ be the maximum length of a dependency in $\Sigma$;
\STATE let $h$ be the number of dependencies in $\Sigma$;
\STATE \textbf{if} there exists an image $M$ of $Q$ in $\D$ such that
\STATE \quad there exists no $\D'\subseteq\D$ such that
\STATE \qquad (a) $|\D'|\leq k^{h+1}$ and
\STATE \qquad (b) $\fc(\D',\D,\Sigma)$ is consistent with $\Sigma$ and
\STATE \qquad (c) $\fc(\D'\cup M,\D,\Sigma)$ is inconsistent with $\Sigma$
\STATE \textbf{then return} true
\STATE \textbf{else return} false
\end{algorithmic}
\end{algorithm}

\iflong We now show the correctness of the algorithm \algacyclicffk. \fi

\begin{theorem}
\label{thm:algorithm-acyclic-ffk-correct}
Let $\Sigma$ be a set of acyclic dependencies, let $\D$ be a database such that $\Sigma$ is \ffk for $\D$, and $Q$ be a BUCQ. $\SD$ \irs-entails $Q$ iff the algorithm \algacyclicffk$(\Sigma,\D,Q)$ returns true.
\end{theorem}
\iflong 
\begin{proof}
First, suppose $\SD$ \irs-entails $Q$. Then, there exists an image $M$ of $Q$ in $\D$ that belongs to all the repairs of $\SD$.
Now, suppose the algorithm returns false, and let $M$ and $\D'$ satisfy the three conditions (a), (b), (c) of the algorithm.
Since condition (b) holds, by Proposition~\ref{pro:wc-fc} $\D'$ is weakly consistent with $\SD$, thus $\D'\subseteq\R$ for some repair $\R$ of $\SD$. And, since by hypothesis $M$ is contained in each repair, we have that $\D'\cup M\subseteq\R$.
Since $\R$ is consistent with $\Sigma$ and contains $\D'\cup M$, it follows that $\D'\cup M$ is weakly consistent with $\Sigma$ too, thus by Proposition~\ref{pro:wc-fc} condition (c) is false, contradicting the above hypothesis. Consequently, the algorithm returns true.

Conversely, suppose the algorithm returns true and let $M$, $\D'$ satisfy conditions (a), (b), (c) of the algorithm.
Now suppose there exists a repair $\R$ of $\SD$ such that $M\not\subseteq\R$. This implies that $\R\cup M$ is not weakly consistent with $\SD$, therefore by Proposition~\ref{pro:wc-fc} $\fc(\R\cup M,\D,\Sigma)$ is inconsistent with $\Sigma$, hence by Proposition~\ref{pro:consistency} 
there exists a dependency in $\tau\in\Sigma$ and an instantiation $\sigma$ of $\body(\tau)$ in $\fc(\R\cup M,\D,\Sigma)$ such that the BUCQ $\sigma(\head(\tau))$ has no image in $\D$. Let $V$ be the set $\{\sigma(\alpha)\mid\alpha\in\predatoms(\body(\tau))\}$. Observe that $V$ contains at most $k$ facts, where $k$ is the maximum number of predicate atoms occurring in the body of a dependency in $\Sigma$.

Furthermore, since $\Sigma$ is acyclic, the presence of a fact in $\fc(\R\cup M,\D,\Sigma)$ depends on at most $k^h$ facts (where $h$ is the number of dependencies in $\Sigma$), therefore there exists a set of facts $\R'\subseteq\R$ such that $|\R'|\leq k^{h+1}$ and $\fc(\R'\cup M,\Sigma)$ contains $V$, causing $\fc(\R'\cup M,\Sigma)$ to be inconsistent with $\Sigma$.
But this implies that $\R'$ satisfies all the conditions of the set $\D'$ in the algorithm, which therefore returns false, thus contradicting the hypothesis. Hence, every repair of $\SD$ contains $M$, which implies that $Q$ is \irs-entailed by $\SD$.
\end{proof}
\fi 

\iflong 
We now show that the Algorithm \algacyclicffk can be turned into a query rewriting technique:
given a set of acyclic dependencies $\Sigma$ and a BUCQ $Q$, it is possible to compute a first-order query $\phi(Q,\Sigma)$ such that, for every database $\D$ such that $\Sigma$ is \ffk for $\D$, $Q$ is \irs-entailed by $\SD$ iff $\phi(q,\Sigma)$ is evaluated to $\true$ over $\D$.
That is, we can actually express the conditions of the algorithm \algacyclicffk in terms of an FO sentence, exploiting the formula $\weaklycons(\Sigma)$ defined in Section~\ref{sec:wc} to encode the weak consistency problem for acyclic \ffk dependencies.

\fi 
Given a set of atoms $\A$ (possibly with variables), we define $\weaklycons(\A,\Sigma)$ as the formula obtained from $\weaklycons(\Sigma)$ by replacing every atom $\aux(p(t))$ (with $m$ arguments) with the subformula $\bigvee_{p(x)\in\A} (p(x) \wedge \bigwedge_{i=1}^m t_i=x_i))$.



The following property is immediately implied by Theorem~\ref{thm:weaklycons} and the definition of $\weaklycons(\A,\Sigma)$.

\begin{lemma}
\label{lem-weakly-acyclic-fo-rewriting}
Let $\A$ be a set of atoms, let $y$ be a tuple of the variables occurring in $\A$, and let $\sigma$ be an instantiation of $\bigwedge_{\alpha\in\A}\alpha$ in $\D$. Then, the sentence $\sigma(\weaklycons(\A,\Sigma))$ evaluates to true over $\D$ iff $\{\sigma(\alpha)\mid\alpha\in\A\}$ is weakly consistent with $\SD$.
\end{lemma}

Now we want to simulate the check of weak consistency of any possible subset $\D'$ of $\D$ of size at most $k^{h+1}$ of algorithm \algacyclicffk using (the conjunction of a finite number of occurrences of) the above formula $\weaklycons(\A,\Sigma)$. To do so, we need sets of atoms $\A$ whose instantiations in $\D$ pick the above subset $\D'$ and whose weak consistency is then evaluated by the subformula $\weaklycons(\A,\Sigma)$.

Therefore, we now consider sets of atoms $\A$ containing at most $k^{h+1}$ atoms, 
and such that all the arguments of the atoms in $\A$ are variables and every variable occurs only once in $\A$. If $\ell$ is the number of predicates in $\Sigma$, the number of all such sets of atoms (up to renaming of variables) is 
not greater than $2 \ell^{k^{h+1}}$.
Let $\atomsets(\Sigma)$ be the set of all possible such sets of atoms.

Then, given a BCQ $q=\exists x\, (\alpha_1\wedge\ldots\wedge\alpha_k\wedge\ineq_1\wedge\ldots\wedge\ineq_\ell)$, we define the following sentence $\queryrewrite(q,\Sigma)$:
\begin{equation}\label{eqn:rewr-ir-acyclic-ffk}
\begin{array}{r@{}l}
\exists x\,\Big(&\alpha_1\wedge\ldots\wedge\alpha_k\wedge\ineq_1\wedge\ldots\wedge\ineq_\ell\wedge\\

&\quad\;\displaystyle
\bigwedge_{\A\in\atomsets(\Sigma)} 
\forall y\,
\Big(\Big(\bigwedge_{\beta\in\A} \beta\Big) \rightarrow \big(\weaklycons(\A,\Sigma) \ra 
\weaklycons(\A\cup\predatoms(q),\Sigma)\big)\Big)\Big)
\end{array}
\end{equation}
where $y$ is the sequence of all the variables occurring in $\A$.


Moreover, given a BUCQ $Q$, we define 
\[
\displaystyle
\queryrewrite(Q,\Sigma)=\bigvee_{q\in\cq(Q)}\queryrewrite(q,\Sigma)
\]
\smallskip

From Algorithm \algacyclicffk and Lemma~\ref{lem-weakly-acyclic-fo-rewriting} it follows that:

\begin{theorem}
\label{thm:fo-rewr-ir-acyclic-ffk-correct}
Let $\Sigma$ be a set of acyclic dependencies and $Q$ be a BUCQ. For every database $\D$ such that each dependency of $\Sigma$ is \ffk for $\D$, $Q$ is \irs-entailed by $\SD$ iff the sentence $\queryrewrite(Q,\Sigma)$ is true in $\D$.
\end{theorem}

\begin{corollary}\label{cor:ir-ub-acyclic-ffk}
Let $\Sigma$ be a set of acyclic dependencies, let $\D$ be a database such that $\Sigma$ is \ffk for $\D$, and $Q$ be a BUCQ. Deciding whether $Q$ is \irs-entailed by $\SD$ is $\aczero$ \wrt data complexity.
\end{corollary}

\begin{example}\label{ex:ir-acyclic-ffk}
    Let $\D$ and $\Sigma$ be as in Example~\ref{ex:wc-acyclic-ffk}.
    It is straightforward to see that $\D$ is inconsistent with $\constr$.
    \iflong
    There are two minimal ways of solving such an inconsistency. In both cases, $R(e,e)$ must be deleted because of the first dependency. Furthermore, due to the second dependency, we must choose whether to remove the fact $P(e,f)$ or $T(e,g)$.
    Thus, the
    \fi
    set of repairs of $\SD$ is $\reps(\D)=\{ \D\setminus\{T(e,g),R(e,e)\}, \D\setminus\{P(e,f),R(e,e)\} \}$ and their intersection is 
    $\intrep(\D)=\{ P(a,b), T(a,c), R(a,d) \}$.
    
    Let us now consider the query $q=\exists z'\, T(e,z')$. We have that $\intrep(\D)\not\models q$ (in fact, $T(e,z')$ has no image in $\intrep(\D)$). By Theorem~\ref{thm:fo-rewr-ir-acyclic-ffk-correct}, this can 
    be verified by evaluating $\queryrewrite(q,\Sigma)$ over $\D$.
    To do this, let us consider the instantiation of Equation~\ref{eqn:rewr-ir-acyclic-ffk} with inputs $q$ and $\constr$, one of the conjuncts of which (for $\A=\{P(x',y')\}$) is:
    \[
    \begin{array}{r@{}l}
        \phi=\forall x',y'\,\big(P(x',y') \ra (&\neg\weaklycons(\{ P(x',y') \},\constr) \vee \\
         &\weaklycons(\{ P(x',y'), T(e,z') \},\constr))\big)
    \end{array}
    \]
    Now, considering the sentence $\weaklycons(\constr)$ provided in Example~\ref{ex:wc-acyclic-ffk},
    we can compute the two disjuncts of the consequent of $\phi$ as follows:
    \[
    \begin{array}{l}
    \weaklycons(\{ P(x',y') \},\constr) = \true \\
    \weaklycons(\{ P(x',y'), T(e,z') \},\constr) = \\
    \qquad \forall x,y,z\,\big((P(x',y') \wedge x=x' \wedge y=y' \wedge T(e,z') \wedge x=e \wedge z=z') \ra \bot\big) \wedge\\
    \qquad \forall v\,\big(\big(\exists y,z\,(P(v,y) \wedge T(v,z) \wedge R(v,v) \wedge z \ne v \wedge P(x',y') \wedge v=x' \wedge y=y' \wedge\\
    \qquad\qquad\qquad\quad T(e,z') \wedge v=e \wedge z=z')\big) \ra \bot\big)
    \end{array}
    \]
    (
    $\weaklycons(\{ P(x',y') \},\constr)$ turns out to be a tautology because, since the only atom in the input set has predicate $P$, the atoms with predicate $T^\aux$ and $R^\aux$ in $\weaklycons(\constr)$ must be replaced with $\false$).
    It is now 
    easy to check that both $\neg\weaklycons(\{ P(x',y') \},\Sigma)$ and $\weaklycons(\{ P(x',y'), T(e,z') \},\Sigma)$ are false if evaluated over $\D$. Thus, the evaluation of $\phi$ (and, by consequence, of the whole formula $\queryrewrite(q,\Sigma)$) over $\D$ is $\false$.
\end{example}

\iflong
\subsection{Acyclic linear dependencies}

For the case of linear dependencies, we first show the following property, which is an immediate consequence of Lemma~\ref{lem:wc-linear}.

\begin{lemma}
\label{lem:qe-linear}
Let $\Sigma$ be a set of linear dependencies, let $\D$ be a database, and let $Q$ be a BUCQ. Then, $Q$ is \ars-entailed (or, equivalently, \irs-entailed) by $\SD$ iff there exists an image $M$ of $Q$ in $\D$ such that, for each $\alpha\in M$, $\{\alpha\}$ is weakly-consistent with $\SD$.
\end{lemma}

Based on the above property, we can exploit the formula $\weaklyconsacycliclinear(\alpha,\Sigma)$ defined in Section~\ref{sec:wc} to define an FO sentence that is able to decide query entailment for acyclic linear dependencies.
 
Given a BCQ $q$ of the form $\exists x\, (\alpha_1\wedge\ldots\wedge\alpha_k\wedge\ineq_1\wedge\ldots\wedge\ineq_h)$, let $\queryentails(q,\Sigma)$ be the following FO sentence:
\[
\exists x\, 
\Big(\alpha_1\wedge\ldots\wedge\alpha_k\wedge\ineq_1\wedge\ldots\wedge\ineq_h\wedge
\bigwedge_{i=1}^k \weaklyconsacycliclinear(\alpha_i,\Sigma)\Big)
\]
Moreover, given a BUCQ $Q$, we define 
\[
\queryentails(Q,\Sigma)=\bigvee_{q\in\cq(Q)}\queryentails(q,\Sigma)
\]

The following theorem is an immediate consequence of the above definition of $\queryentails(Q,\Sigma)$, Lemma~\ref{lem:qe-linear} and Lemma~\ref{lem:weaklyconsacycliclinear}.

\begin{theorem}
\label{thm:fo-rewriting-acyclic-linear-correct}
Let $\Sigma$ be a set of acyclic linear dependencies, let $\D$ be a database, and let $Q$ be a BUCQ. Then, $Q$ is \ars-entailed (or equivalently \irs-entailed) by $\SD$ iff $\D\models\queryentails(Q,\Sigma)$.
\end{theorem}

The above theorem immediately implies the following upper bound.

\begin{theorem}
\label{thm:qe-ub-acyclic-linear}
BUCQ entailment in the case of acyclic linear dependencies is $\aczero$ \wrt data complexity.
\end{theorem}

\begin{example}\label{ex:qe-acyclic-linear}
    Let $\D$ and $\constr$ be as in Example~\ref{ex:wc-acyclic-linear}.
    It is straightforward to see that $\D$ is inconsistent with $\constr$.
    Not surprisingly, as we are under the conditions of Proposition~\ref{pro:cq-entailment-correspondence}, there is only one minimal way of solving such an inconsistency, i.e. deleting $T(b,c)$ and, consequently, $P(a,b)$.
    Thus, we have that
    $\reps(\D) = \{ \intrep(\D) \} = \{ \{ T(a,d), T(a,e), R(a,d,b) \} \}$.
    
    Let us now consider the queries $q_1=\exists x,y,z\, (T(x,y) \wedge T(x,z) \wedge y \ne z)$ and $q_2=\exists x,y,z\, (P(x,y) \wedge R(x,y,z))$. We have that
    $\queryentails(q_1,\constr)=\exists x,y,z\, (T(x,y) \wedge T(x,z) \wedge y \ne z \wedge \exists v,w\,R(x,v,w))$ and
    $\queryentails(q_2,\constr)=\exists x,y,z\, (P(x,y) \wedge R(x,y,z) \wedge (\exists z\,(T(y,z) \wedge y\ne z \wedge \exists v,w\,R(y,v,w))))$.
    One can verify that $\intrep(\D)\models q_1$ as well as $\D\models\queryentails(q_1,\constr)$ and that $\intrep(\D)\not\models q_2$ as well as $\D\not\models\queryentails(q_2,\constr)$.
\end{example}

\fi


%% file: 9-conclusions.tex
\section{Conclusions}
\label{sec:conclusions}



\iflong
In this paper, we have studied CQA in the context of disjunctive embedded dependencies with inequalities, a very expressive language for schema constraints. We have studied four decision problems related to the notion of database repair under tuple-deletion semantics for the whole class of disjunctive dependencies and for the linear, acyclic and forward-deterministic subclasses. We have shown tractability \wrt data complexity of the four decision problems for several of such classes of dependencies.
\fi

In our opinion, one of the most interesting future research directions of this work is to study the complexity of the problems considered in this paper under semantics different from the tuple-deletion semantics, in particular, the so-called \emph{symmetric-difference} semantics.

We are also interested in analyzing the complexity of both the \ars-entailment and the \irs-entailment problems for queries more expressive than BUCQs.
For instance, it is straightforward to verify that Proposition~\ref{pro:algorithm-ars-correct}, Proposition~\ref{pro:algorithm-irs-correct}, Theorem~\ref{thm:ar-ub-general-and-ffk}, Theorem~\ref{thm:ir-ub-general-and-ffk} and Theorem~\ref{thm:ar-ub-acyclic} also hold when the query is an arbitrary (domain-independent) first-order sentence.

Another very important aspect is the development of practical algorithms for these problems. We are especially interested in the cases for which the data complexity of the problem is in the class $\aczero$. We have shown that such problems can be solved by the evaluation of a first-order sentence over the database: this could be a starting point towards the development of practical algorithms for such cases.

Finally, CQA has interesting connections with the problem of \emph{controlled query evaluation} \cite{LRS19}, i.e.\ the problem of evaluating queries on a database (or knowledge base) in the presence of a logical specification of a privacy policy that should not be violated by the query answers. We are very interested in investigating the consequences of our results for such a problem.

%% file: appendix.tex
\clearpage
\setcounter{secnumdepth}{0}
\section{Appendix}

%
%

\noindent\textbf{Related work:}

Consistent Query Answering was originally proposed for relational databases in \cite{ABC99}, which introduced the notions of repairs and consistent answers.
Ever since, many works studied the complexity of CQA considering different kinds of integrity constraints and adopting different repairing approaches.
In particular, as said the introduction, in the following we consider repairs based on tuple-deletion, with a specific focus on the problems of repair checking and (conjunctive) query entailment.

For what concerns repair checking, the most relevant works related to our investigation are \cite{CM05,AK09,SC10,GO10,CFK12}, which deeply studied the problem for many classes of dependencies.

Also the problem of finding consistent answers for a given query under tuple-deletion semantics has been extensively studied by \cite{CM05} and \cite{CFK12}, with a particular focus on (some classes of) tuple-generating dependencies, providing complexity results ranging from PTIME to undecidability.
Special attention should also be paid to \cite{FM05} and \cite{KW17}, which proposed first-order rewritable techniques for solving the CQA problem, though limiting the set of integrity constraints to (primary) key dependencies and making some assumptions on the user query.

In the mentioned works, all of which are best reviewed in \cite{B19}, the most common semantics adopted corresponds to the one we will call \ars.
The \irs\ semantics, instead, was previously studied in the context of ontologies with the name of IAR ("intersection of ABox repairs"), and investigated for multiple DL languages in \cite{LLRRS15}.
In particular, the latter work proved conjunctive query entailment under IAR semantics to be first-order rewritable for the language $DL\text{-}Lite_{R,den}$.
Afterward, \cite{B12} proposed a new semantics named ICR ("intersection of closed repairs"), which IAR is a sound approximation of, showing its first-order expressibility for simple ontologies.
However, such results are not straightforwardly transposable under closed-world assumption, which is our case study.

\bigskip


\noindent\textbf{Example on \ars-entailment and \irs-entailment (Section ~\ref{sec:repairs})}:

The following example shows that, in general, the two semantics are different.
\begin{example}\label{ex:semantics-difference}
    Let
    $\constr=\{
        \forall x,y,z\, (P(x,y) \wedge P(x,z) \wedge y \ne z \ra \bot),
        \forall x\, (T(x) \ra \exists y\, (P(y,x)))
    \}$,
    $\D=\{ P(c,a), P(c,b), P(d,c), T(a), \allowbreak T(b)\}$.
    It's easy to see that the first dependency of $\constr$ is not satisfied by $\D$.
    There are two minimal ways of solving the inconsistency between $\D$ and $\constr$ by means of tuple-deletion. The first one consists in deleting the fact $P(c,a)$ and, due to the second dependency, also the fact $T(a)$. Analogously, the second way consists in deleting both $P(c,b)$ and $T(b)$.
    Thus,
    $\reps(\D)=\{ \{P(c,a), P(d,c), T(a)\}, \{P(c,b), P(d,c), T(b)\} \}$ and
    $\intrep(\D)=\{ P(d,c) \}$.
    Let us now consider the sentence $\phi=\exists x\, P(c,x)$:
    $\phi$ is entailed by all the repairs, but not by their intersection, i.e.,  is \ars-entailed but not \irs-entailed by $\SD$.
\end{example}

\bigskip

\noindent\textbf{Proof of Theorem~\ref{thm:wc-lb-linear-ffk}}:

We prove the thesis by showing a reduction from STCON (the reachability problem on directed graphs).
Let $\Sigma$ be the following set of linear dependencies:
\[
\begin{array}{l}
\forall x,y,z\,(\succ(x,y,z)\rightarrow \vert(y)) \\
\forall x,y,z\,(\succ(x,y,z)\rightarrow \exists w\,(\succ(x,z,w))) \\
\forall x\,(\vert(x)\rightarrow \exists y\,(\succ(x,0,y)))
\end{array}
\]

Now let $G=\tup{V,E}$ be a directed graph ($V$ is the set of vertices of $G$ and $E$ is the set of edges of $G$) and let $s,t\in V$. W.l.o.g.\ we assume that $G$ is represented through an adjacency list and that $0\not\in V$.
We define the following set of facts $\D$: 
\[
\begin{array}{l}
\{ \vert(a) \mid a\in V \wedge a\neq t \} \;\cup \\
\{ \succ(a,0,0) \mid a\in V \wedge \textit{ a has no successors in } G \} \;\cup \\
\{ \succ(a,0,b_1),\succ(a,b_1,b_2),\ldots,\succ(a,b_{h-1},b_h),\succ(a,b_h,0) \mid \\
\quad\quad a\in V \wedge\tup{b_1,\ldots,b_h} \textit{ is the adjacency list of } a \} 
\end{array}
\]
Observe that $\Sigma$ is a set of \ffk dependencies for $\D$ (in particular, for every $x$ there is at most one fact in $\D$ of the form $\succ(x,0,\cdot)$, and for every $x,y$ there is at most one fact in $\D$ of the form $\succ(x,y,\cdot)$).
Finally, let $\D'=\{\vert(s)\}$.

It is possible to verify that there exists a path in $G$ from $s$ to $t$ iff $\D'$ is not weakly consistent with $\SD$. The key point is that $\Sigma$ is such that: $(i)$ all the incoming edges of $t$, i.e., all the facts of the form $\succ(\cdot,t,\cdot)$, must be deleted in all repairs due to the first dependency and the absence of $\vert(t)$ in $\D$; $(ii)$ due to the second dependency, the elimination of one edge $(a,b)$ represented by the fact $\succ(a,b,n)$ implies the elimination of all the outgoing edges of $a$, i.e.\ the elimination of all the facts of the form $\succ(a,\cdot,\cdot)$. This in turn implies, by the third dependency, the elimination of the fact $\vert(a)$. Consequently, for every vertex $a$, $\vert(a)$ belongs to the only repair of $\SD$ iff $t$ is not reachable from $a$ in $G$.
\qed

\bigskip

\noindent\textbf{Proof of Theorem~\ref{thm:wc-lb-ffk-and-linear}}:

First, we prove thesis $(i)$ by reduction from HORN 3-SAT (a well-known PTIME-complete problem). We define the following set of \ffk (actually, full) dependencies $\Sigma$:
\[
\begin{array}{l}
\forall x\,(C(0,0,x)\rightarrow A(x)) \\
\forall x,y\,(C(x,0,y)\wedge A(x)\rightarrow A(y)) \\
\forall x,y,z\,(C(x,y,z)\wedge A(x)\wedge A(y)\rightarrow A(z)) \\
\forall x\,(C_f(x,0,0)\wedge A(x)\rightarrow \bot) \\
\forall x,y\,(C_f(x,y,0)\wedge A(x)\wedge A(y)\rightarrow \bot) \\
\forall x,y,z\,(C_f(x,y,z)\wedge A(x)\wedge A(y)\wedge A(z)\rightarrow \bot)
\end{array}
\]
\
Then, given a Horn 3-CNF $\phi$, we define $\D'$ as the set containing the following facts:
\begin{itemize}
\item
$C(0,0,a)$ for each clause of the form $a$ in $\phi$;
\item
$C(a,0,b)$ for each clause of the form $\neg a \vee b$ in $\phi$;
\item
$C(a,b,c)$ for each clause of the form $\neg a \vee \neg b \vee c$ in $\phi$;
\item
$C_f(a,0,0)$ for each clause of the form $\neg a$ in $\phi$;
\item
$C_f(a,b,0)$ for each clause of the form $\neg a \vee \neg b$ in $\phi$;
\item
$C_f(a,b,c)$ for each clause of the form $\neg a \vee \neg b \vee \neg c$ in $\phi$.
\end{itemize}
Moreover, let $V$ be the set of propositional variables occurring in $\phi$. We define the set of facts $\D''=\{ A(a) \mid a\in V\}$. Finally, let $\D=\D'\cup\D''$.

It can be shown that the models of $\phi$ are in 1-to-1 correspondence with the repairs of $\SD$ containing $\D'$. Consequently, $\phi$ is satisfiable iff there exists a repair of $\SD$ containing $\D'$, which implies that $\phi$ is satisfiable iff $\D'$ is weakly consistent with $\SD$.

\medskip

Then, we prove thesis $(ii)$ by showing that the HORN SAT problem can be reduced to (non-)weak consistency.

Given a set of $n$ ground Horn rules (we represent rules without head with an extra variable $\false$ and assume w.l.o.g.\ that there is at least one rule without head whose id is $r_1$), we represent a rule $r_i$ of the form $a \leftarrow b_1,\ldots,b_k$ in the database $\D$ as follows:
$H(r_i,a,1,1)$ (if there are multiple rules $r_{i_1},\ldots,r_{i_h}$ having $a$ in the head, we write $H(r_{i_1},a,1,2), H(r_{i_1},a,2,3), \ldots, H(r_{i_h},a,h,1)$), $B(r_i,b_1), \ldots, B(r_i,b_k)$.

We define the following set of dependencies $\Sigma$:
\[
\begin{array}{l}
\forall x,y,z,w \,(H(x,y,z,w)\rightarrow \exists v \, B(x,v)) \\
\forall x,y,z,w \,(H(x,y,z,w)\rightarrow \exists v,t \, H(v,y,w,t)) \\
\forall x,y\,(B(x,y)\rightarrow \exists z,w,v \, H(z,y,w,v))
\end{array}
\]
where:
the first dependency implies that if for a rule there are no more body atoms, the head atom of that rule is deleted;
the second dependency implies that if a head atom is deleted, all the head atoms with that variable are deleted;
the third dependency implies that if there are no head atoms for a variable, all the body atoms for that variable are deleted.

Therefore, it can be verified that the Horn formula is unsatisfiable iff the fact $H(r_1,\false,1,2)$ belongs to the only repair of $\SD$, i.e.\ iff the set  $\{H(r_1,\false,1,2)\}$ is weakly consistent with $\SD$.
\qed 

\bigskip

\noindent\textbf{Proof of Theorem~\ref{thm:wc-lb-acyclic}}:

The proof is obtained through a reduction from the 3-CNF problem. Let $\Sigma$ be the following set of acyclic dependencies:
\[
\begin{array}{l}
\forall x\,(R(x)\rightarrow \exists y\, V(x,y)) \\
\forall w,x_1,x_2,x_3,y_1,y_2,y_3\,
(C(w,x_1,y_1,x_2,y_2,x_3,y_3)\wedge \\ \quad V(x_1,y_1)\wedge
V(x_2,y_2)\wedge V(x_3,y_3)\rightarrow \bot)
\end{array}
\]
Now, let $\phi=\bigwedge_1^n(\ell_i^1\vee \ell_i^2\vee\ell_i^3)$ be a 3-CNF formula, $A$ be the set of propositional variables occurring in $\phi$, and $\D$ be the following set of facts:
\[
\begin{array}{l}
\{ C(i,\propvar(\ell_i^1),\ntv(\ell_i^1),\propvar(\ell_i^2),\ntv(\ell_i^2),\propvar(\ell_i^3),\ntv(\ell_i^3)) \mid \\\quad 1\leq i\leq n \}
\cup \{ R(a), V(a,0), V(a,1) \mid a\in A \}
\end{array}
\]
where each $\propvar(\ell_i^j)$ is the propositional variable appearing in the literal $\ell_i^j$ and each $\ntv(\ell_i^j)$ is 1 if the literal $\ell_i^j$ is negated and 0 otherwise.
Finally, let $\D'$ be the following set of facts:
\[
\begin{array}{l}
\{ C(i,\propvar(\ell_i^1),\ntv(\ell_i^1),\propvar(\ell_i^2),\ntv(\ell_i^2),\propvar(\ell_i^3),\ntv(\ell_i^3)) \mid \\\quad 1\leq i\leq n \}
\cup \{ R(a) \mid a\in A \}
\end{array}\]
(note that both $\D$ and $\D'$ are inconsistent with $\Sigma$).

It is now easy to verify that $\D'$ is weakly consistent with $\SD$ iff $\phi$ is satisfiable.
Let us first consider the case in which $\phi$ is satisfiable. Then there exists a guess of the variables of $\phi$ such that each clause of $\phi$ is satisfied.
Such guess can be represented with a set $\D''$ of facts with predicate $V$ corresponding to the images of the head of the first dependency.
Moreover, since all the clauses of $\phi$ must be satisfied, each of them must contain at least one variable having a truth value equal to 0 if the literal is negated and 1 otherwise.
Therefore, the set $\D'\cup \D''$ does not contain any image of the body of the second dependency. Thus $\D'\cup\D''$ is a repair of $\SD$ and, consequently, $\D'$ is weakly consistent with $\SD$.

Conversely, if $\phi$ is unsatisfiable, then notice that every repair $\D'\cup\D''$ containing $\D'$ should be such that $\D''$ contains a fact $V(a,\cdot)$ for every propositional variable $a$ (otherwise $\D'\cup\D''$ would be inconsistent with the first dependency). On the other hand, since $\phi$ is unsatisfiable, each possible set $\D''$ representing a guess of the truth values of the variables of $\phi$ is such that $\D'\cup\D''$ is inconsistent with the second dependency. Therefore, $\D'$ is not weakly consistent with $\SD$.
\qed 

\bigskip

\noindent\textbf{Proof of Theorem~\ref{thm:rc-lb-linear-ffk}}:

We prove the thesis by showing a reduction from STCON (the reachability problem on directed graphs).
Let $\Sigma$ be the following set of linear dependencies:
\[
\begin{array}{l}
\forall x,y,z\,(\succ(x,y,z)\rightarrow \vert(y)) \\
\forall x,y,z\,(\succ(x,y,z)\rightarrow \exists w\,(\succ(x,z,w))) \\
\forall x\,(\vert(x)\rightarrow \exists y\,(\succ(x,0,y))) \\
\forall x\,(\initial(x)\rightarrow \vert(x)) \\
\forall x\,(\vert(x)\rightarrow \exists y\,\initial(y)) \\
\forall x,y,z\,(\succ(x,y,z)\rightarrow \exists w\,\initial(w)) 
\end{array}
\]

Now, let $G=\tup{V,E}$ be a directed graph ($V$ is the set of vertices of $G$ and $E$ is the set of edges of $G$) and let $s,t\in V$. W.l.o.g.\ we assume that $G$ is represented through an adjacency list and that $0\not\in V$.
We define the following set of facts $\D$: 
\[
\begin{array}{l}
\{ \initial(s) \} \cup
\{ \vert(a) \mid a\in V \wedge a\neq t \} \;\cup \\
\{ \succ(a,0,0) \mid a\in V \wedge \textit{ a has no successors in } G \} \;\cup \\
\{ \succ(a,0,b_1),\succ(a,b_1,b_2),\ldots,\succ(a,b_{h-1},b_h),\succ(a,b_h,0) \mid \\
\quad\quad a\in V \wedge\tup{b_1,\ldots,b_h} \textit{ is the adjacency list of } a \} 
\end{array}
\]
Observe that $\Sigma$ is a set of \ffk dependencies for $\D$ (in particular, for every $x$ there is at most one fact in $\D$ of the form $\succ(x,0,\cdot)$, and for every $x,y$ there is at most one fact in $\D$ of the form $\succ(x,y,\cdot)$).
Finally, let $\D'=\emptyset$.

It is possible to verify that there exists a path in $G$ from $s$ to $t$ iff $\D'$ is a repair of $\SD$. The key point is that $\Sigma$ is such that: $(i)$ all the incoming edges of $t$, i.e., all the facts of the form $\succ(\cdot,t,\cdot)$, must be deleted in all repairs due to the first dependency and the absence of $\vert(t)$ in $\D$; $(ii)$ due to the second dependency, the elimination of one edge $(a,b)$ represented by the fact $\succ(a,b,n)$ implies the elimination of all the outgoing edges of $a$, i.e.\ the elimination of all the facts of the form $\succ(a,\cdot,\cdot)$. This in turn implies, by the third dependency, the elimination of the fact $\vert(a)$.
Therefore, if $t$ is reachable from $s$ in $G$, then $\vert(s)$ cannot belong to the repair of $\SD$, which by the fourth dependency implies that $\initial(s)$ cannot belong to the repair of $\SD$, which by the fifth and sixth dependency implies that no other facts of $\D$ cannot belong to the repair of $\SD$, i.e.\ the repair of $\SD$ is the empty set.
\qed 

\bigskip

\noindent\textbf{Proof of Theorem~\ref{thm:ic-lb-linear-and-linear-ffk-and-acyclic}}:

The proof of thesis $(i)$ is obtained through a reduction from 3-CNF.
Let $\Sigma$ be the following set of acyclic dependencies:
\[
\begin{array}{l}
\forall x\,(V(x,1)\wedge V(x,0)\rightarrow \bot) \\
\forall x,y_1,y_2,y_3,z_1,z_2,z_3\,
(C_1(x,y_1,z_1,y_2,z_2,y_3,z_3)\wedge\\
\quad V(y_1,z_1)\wedge V(y_2,z_2)\wedge V(y_3,z_3) \rightarrow \bot) \\
\ldots \\
\forall x,y_1,y_2,y_3,z_1,z_2,z_3\,
(C_7(x,y_1,z_1,y_2,z_2,y_3,z_3)\wedge\\
\quad V(y_1,z_1)\wedge V(y_2,z_2)\wedge V(y_3,z_3) \rightarrow \bot) \\
\forall x,y_1,y_2,y_3,y_4,y_5,y_6\,\\
\quad
(C_1(x,y_1,y_2,y_3,y_4,y_5,y_6) \rightarrow \\
\qquad
\exists w_1,w_2,w_3,w_4,w_5,w_6\: C_2(x,w_1,w_2,w_3,w_4,w_5,w_6)) \\
\ldots \\
\forall x,y_1,y_2,y_3,y_4,y_5,y_6\,\\
\quad
(C_6(x,y_1,y_2,y_3,y_4,y_5,y_6) \rightarrow \\
\qquad
\exists w_1,w_2,w_3,w_4,w_5,w_6\: C_7(x,w_1,w_2,w_3,w_4,w_5,w_6)) \\
U \rightarrow \exists x,y_1,y_2,y_3,y_4,y_5,y_6\: C_1(x,y_1,y_2,y_3,y_4,y_5,y_6)
\end{array}
\]
Then, let $\D$ be the following database that contains:
$V(a,0),V(a,1)$ for each variable $a$ occurring in $\phi$;
7 facts $C_1, \ldots, C_7$ for each clause of $\phi$ (each such fact represents an evaluation of the three variables of the clause that make the clause true);
and the fact $U$.

E.g. if the $i$-th clause is $\neg a \vee \neg b \vee c$, then $\D$ contains the facts
\[
\begin{array}{l}
C_1(i,a,0,b,0,c,0),
C_2(i,a,0,b,0,c,1),
C_3(i,a,0,b,1,c,0),\\
C_4(i,a,0,b,1,c,1),
C_5(i,a,1,b,0,c,0),
C_6(i,a,1,b,0,c,1),\\
C_7(i,a,1,b,1,c,1)
\end{array}
\]

We prove that the fact $U$ belongs to all the repairs of $\SD$ iff $\phi$ is unsatisfiable.

In fact, if $\phi$ is unsatisfiable, then there is no repair of $\SD$ without any fact of $C_1$, consequently $U$ belongs to all the repairs of $\SD$.
Conversely, if $\phi$ is satisfiable, then there is a repair (where the extension of $V$ corresponds to the interpretation satisfying $\phi$) that for each clause does not contain at least one of the 7 facts $C_1,\ldots,C_7$ representing the clause. Due to the dependencies between $C_i$ and $C_{i+1}$, this implies that there is a repair that for each clause does not contain any fact $C_1$, and therefore (due to the last dependency) it does not contain the fact $U$.

\medskip 

The proof of thesis $(ii)$ immediately follows from the reduction shown in the proof of thesis $(i)$,
which already shows that, given $\Sigma$ and $\D$ defined as in that proof, the Horn formula is unsatisfiable iff the fact $H(r_1,\false,1,2)$ belongs to the only repair of $\SD$.

\medskip 

The proof of thesis $(iii)$ immediately follows from the reduction shown in the proof of Theorem~\ref{thm:rc-lb-linear-ffk}: it is immediate to see that, in that proof, there exists a path in $G$ from $s$ to $t$ iff the fact $\initial(s)$ does not belong to the only repair of $\SD$.
\qed 

\bigskip

\noindent\textbf{Proof of Theorem~\ref{thm:ic-lb-ffk}}:

The proof is by reduction from 3-CNF.
We define the following set $\Sigma$ of full dependencies:

\[
\begin{array}{l}
\forall x_1,x_2,x_3,v_1,v_2,v_3,y,z\,\\
\quad 
(S(z)\wedge N(y,z)\wedge C(y,x_1,v_1,x_2,v_2,x_3,v_3)\wedge V(x_1,y_1)\rightarrow S(y)) \\
\forall x_1,x_2,x_3,v_1,v_2,v_3,y,z\,\\
\quad 
(S(z)\wedge N(y,z)\wedge C(y,x_1,v_1,x_2,v_2,x_3,v_3)\wedge V(x_2,y_2)\rightarrow S(y)) \\
\forall x_1,x_2,x_3,v_1,v_2,v_3,y,z\,\\
\quad 
(S(z)\wedge N(y,z)\wedge C(y,x_1,v_1,x_2,v_2,x_3,v_3)\wedge V(x_3,y_3)\rightarrow S(y)) \\
\forall x\,(V(x,0)\wedge V(x,1)\rightarrow u) \\
S(1) \rightarrow u
\end{array}
\]

Given a 3-CNF formula $\phi$, in the database $\D$, we represent every clause of $\phi$ with (at most) 7 facts corresponding to the interpretations of the variables that satisfy the clause: e.g.\ if clause $n$ is $a\vee \neg b \vee c$, we add the facts

\[
\begin{array}{l}
C(n,a,0,b,0,c,0),C(n,a,0,b,0,c,1),C(n,a,0,b,1,c,1),\\
C(n,a,1,b,0,c,0),C(n,a,1,b,0,c,1),C(n,a,1,b,1,c,0),\\
C(n,a,1,b,1,c,1)
\end{array}
\]

Moreover, the database contains the facts $V(p,0),V(p,1)$ for every propositional variable $p$, and the facts 

\[
\begin{array}{r@{}l}
\{ & N(a_1,a_2),...,N(a_{m-1},a_m),N(a_m,a),\\
& S(a_1),\ldots,S(a_m),S(a) \}
\end{array}
\]

We prove that $\phi$ is unsatisfiable iff the fact $S(a)$ belongs to all the repairs of $\SD$.

First, if $\phi$ is satisfiable, then let $P$ be the set of propositional variables occurring in $\phi$, let $I$ be an interpretation (subset of $P$) satisfying $\phi$, and let $\D'$ be the following subset of $\D$:

\[
\begin{array}{r@{}l}
\D' = \D \setminus (
    &\{V(p,0) \mid p\in P\cap I \}\;\cup\\
    &\{ V(p,1) \mid p\in P\setminus I \} \cup \{S(a)\} )
\end{array}
\]

It is immediate to verify that $\D'$ is consistent with $\Sigma$ and that $\D'\cup \{s(a)\}$ is not weakly consistent with $\SD$: in fact, the addition of $S(a)$ creates a sequence of instantiations of the bodies of the first three dependencies of $\Sigma$ that requires (to keep the consistency of the set) to add to $\D'\cup\{s(a)\}$ first the fact $s(a_m)$, then $s(a_{m-1}$, and so on until $s(a_1)$,but this makes this set inconsistent with $\Sigma$, due to the last dependency of $\Sigma$ and the absence of the fact $u$ in $\D$. Consequently, there exists a repair of $\SD$ that does not contain $s(a)$.

On the other hand, if $\phi$ is unsatisfiable, then every guess of the atoms of the $V$ predicate that satisfies the fourth dependency (each such guess corresponds to an interpretation of the propositional variables) is such that the sequence of instantiations of the bodies of the first three dependencies of $\Sigma$ mentioned above, which leads to the need of adding $S(a_1)$ to the set, is blocked by the absence of some fact for $V$. Consequently, the deletion of $S(a_1)$ does not imply the deletion of $S(a)$ and $S(a)$ belongs to all the repairs of $\SD$.
\qed 

\bigskip

\noindent\textbf{Proof of Theorem~\ref{thm:wc-ub-ffk}}:

The proof follows from Proposition~\ref{pro:wc-fc} and from the fact that $\fc(\D'\,\D,\Sigma)$ can be computed in polynomial time w.r.t.\ data complexity through an iterative algorithm that, for every instantiation of the body of a dependency in $\D'$, adds to $\D'$ the corresponding (unique) image of the head of the dependency in $\D$, until a fixpoint is reached (and a fixpoint is reached after at most $n$ additions of new images of dependency heads, if $n$ is the number of facts in $\D$).
\qed 

\bigskip

\noindent\textbf{Proof of Lemma~\ref{lem:belongstofc}}:

To prove the thesis, we need the following property, whose proof is straightforward:

\begin{lemma}
\label{lem:belongstofc-aux}
Let $\Sigma$ be a set of acyclic dependencies, where $\tup{\tau_1,\ldots,\tau_h}$ is a topological order of $\Sigma$. For every fact of the form $p(t)$, $\alpha$ be an atom and let $\sigma$ be a substitution of the variables occurring in $\alpha$ with constants. Then, for every database $\D$ such that $\Sigma$ is \ffk for $\D$, and for every $\D'\subseteq\D$, $p(t)\in\fc(\D',\D,\Sigma)$ iff $p(t)\in\fc(\D',\D,\{\tau_1,\ldots,\tau_j\})$, where $j$ is the largest integer such that the predicate $p$ occurs in the head of $\tau_j$. 
\end{lemma}

Now, in the case when $i=0$, $\belongstofc(\alpha,\Sigma,i)=\aux(\alpha)$, and of course $\fc(\D',\D,\emptyset)=\D'$, and since $\sigma(\alpha)\in\D'$ iff $\D\cup\aux(\D')\models\sigma(\aux(\alpha))$, the thesis follows.

Now suppose that the thesis holds for every $i$ such that $i<\ell$, and consider the case when $i=\ell$.
From the definition of forward closure it follows that $\sigma(\alpha)$ belongs to $\fc(\D',\D,\{\tau_1,\ldots,\tau_i\})$ iff one of these two conditions holds:
\begin{itemize}
\item[$(i)$] $\sigma(\alpha)\in\D'$;
\item[$(ii)$] there exists $\tau_j\in\{\tau_1,\ldots,\tau_i\}$, $q\in\cq(\head(\tau_j))$ and an instantiation $\sigma'$ of $\body(\tau_j)$ such that $\headunify(\sigma(\alpha),q,\sigma')$ is true and $\sigma(\alpha)$ belongs to the (unique) image of $\head(\sigma'(\tau_j))$ in $\D$ and $\bodyp(\sigma'(\tau_j))\subseteq\fc(\D',\D,\{\tau_1,\ldots,\tau_i\})$.
Moreover, Lemma~\ref{lem:belongstofc-aux} implies that $\bodyp(\sigma'(\tau_j))\subseteq\fc(\D',\D,\{\tau_1,\ldots,\tau_i\})$ iff $\bodyp(\sigma'(\tau_j))\subseteq\fc(\D',\D,\{\tau_1,\ldots,\tau_{j-1}\})$.
\end{itemize}

Now, as already mentioned, $\D\cup\aux(\D')\models\sigma(\aux(\alpha))$ iff the above condition $(i)$ holds (notice that $\sigma(\aux(\alpha))$ is the first disjunct of $\sigma(\belongstofc(\alpha,\Sigma,i))$).

Moreover, let $\psi$ be the second disjunct of $\sigma(\belongstofc(\alpha,\Sigma,i))$, i.e.:
\[
\begin{array}{l}
\displaystyle
\bigvee_{\tau_j\in\{\tau_1,\ldots,\tau_i\}}
\bigvee_{\footnotesize\begin{array}{c}q\in\cq(\tau_j) \wedge\\
\headunify(\alpha,q,\sigma)\end{array}} 
\Big(\exists x \big(\sigma(\body(\tau_j))\wedge\sigma(\cnj(q))\wedge \\[2mm]
\displaystyle
\qquad\qquad\qquad\qquad\qquad\qquad\quad
\bigwedge_{\footnotesize\beta\in\bodyp(\sigma(\tau_j))}\belongstofc(\beta,\Sigma,j-1)\big)\Big)
\end{array}
\]
It is immediate to verify that, due to the inductive hypothesis, $\D\cup\aux(\D')\models\psi$ iff the above condition $(ii)$ holds.
Consequently, the thesis follows.
\qed 

\bigskip

\noindent\textbf{Proof of Theorem~\ref{thm:weaklycons}}:

First, observe that 
$\D'$ is weakly consistent with $\SD$ iff 
$\fc(\D',\D,\Sigma)$ is consistent with $\Sigma$, i.e.\ for every dependency $\tau$ and every instantiation $\sigma$ of $\body(\tau)$, there exists $q\in\cq(\head(\tau))$ such that $q$ has an image in $\fc(\D',\D,\Sigma)$.
Then, the thesis follows immediately from the definition of $\weaklycons(\Sigma)$ and 
from Lemma~\ref{lem:belongstofc}.
\qed 

\bigskip

\noindent\textbf{Proof of Theorem~\ref{thm:wc-ub-linear-ffk}}:

Using Lemma~\ref{lem:wc-linear}, we prove the thesis by reducing the problem of deciding the weak consistency of $\{\alpha\}$ with $SD$ to the STCON problem, through the following logspace reduction. Let $G$ be the directed graph such that: $(i)$ there is a vertex $\beta$ for every fact $\beta$ in $\D$; $(ii)$ there is an edge $(\beta,\gamma)$ in $G$ iff there exists a dependency $\tau\in\Sigma$ such that 
there exists an instantiation $\sigma$ of $\body(\tau)$ such that $\bodyp(\sigma(\tau))=\beta$ and $\gamma$ is an image of $\head(\sigma(\tau))$ in $\D$; $(iii)$ there is a further vertex $\bot$ in $G$ and there exists an edge $(\beta,\bot)$ in $G$ iff there exists a dependency $\tau\in\Sigma$ such that 
there exists an instantiation $\sigma$ of $\body(\tau)$ such that $\bodyp(\sigma(\tau))=\beta$ and there exists no image of $\head(\sigma(\tau))$ in $\D$.
It is now easy to verify that $\{\alpha\}$ is not weakly consistent with $\SD$ iff the vertex $\bot$ is reachable from the vertex $\alpha$ in $G$. Then, the thesis follows by Lemma~\ref{lem:wc-linear}.
\qed 

\bigskip

\noindent\textbf{Proof of Theorem~\ref{thm:wc-ub-acyclic-linear}}:

First, we need the following lemma, whose proof easily follows from the definition of $\weaklyconsacycliclinear(\alpha,\Sigma)$:

\begin{lemma}
\label{lem:weaklyconsacycliclinear}
Let $\Sigma$ be a set of acyclic linear dependencies, 
let $\alpha$ be an atom and let $\sigma$ be a substitution of the variables occurring in $\alpha$ with constants. Then, for every database $\D$, 
$\sigma(\alpha)$ is weakly consistent with $\SD$ iff $\sigma(\alpha)\in\D$ and $\D\models\sigma(\weaklyconsacycliclinear(\alpha,\Sigma))$.
\end{lemma}

From Lemma~\ref{lem:weaklyconsacycliclinear}, it immediately follows that $\D'$ is weakly consistent with $\SD$ iff $\D\cup\aux(\D')\models\weaklyconsacycliclinear(\Sigma)$. Now, since $\weaklyconsacycliclinear(\Sigma)$ is an FO sentence, the thesis follows.
\qed 

\bigskip

\noindent\textbf{Proof of Theorem~\ref{thm:acyclic-layers}}:

We prove by induction that, if $\D'$ is not a repair of $\SD$, then there exists $\alpha\in\D'\setminus\D$ such that $\D'\cup\{\alpha\}$ is consistent with $\Sigma$ (the other direction of the proof is straightforward). Let us stratify the set of predicates according to the dependencies (layer 1 = predicates not occurring in the body of any dependency; layer i = predicates occurring only in the body of dependencies such that only predicates of layers lower than i occur in the head). Let $\D_i,\D_i'$ denote respectively the projection of $\D$ and $\D'$ on the predicates of layer $i$. Now:

Base case: if $\D_1\neq\D_1'$, then there exists a fact $\alpha\in\D_1\setminus\D'$ such that $\D'\cup\{\alpha\}$ is consistent with $\Sigma$ (since $\alpha$ cannot generate new instantiations of bodies of dependencies because its predicate does not appear in any body of a dependency);

Inductive case: if $\D_j=\D_j'$ for each $i$ such that $\leq j\leq i$, then for every dependency $\tau$ of the $i$-th layer, every instantiation $\sigma$ of $\body(\tau)$ in $\D$ such that $\D\models\sigma(\head(\tau))$ is also such that $\D'\models\sigma(\head(\tau))$, consequently $\D'\cup\sigma(\body(\tau))$ is consistent with $\Sigma$, hence $\sigma(\body(\tau))\subseteq\D'$ (otherwise $\D'$ would not be a repair of $\SD$). Therefore, if $\alpha \in \D_{i+1}\setminus\D_{i+1}'$, then $\D'\cup\{\alpha\}$ is inconsistent with $\Sigma$.
\qed 

\bigskip

\noindent\textbf{Proof of Theorem~\ref{thm:rc-ub-linear-ffk}}:

Given the hypothesis, it is immediate to verify that $\D'$ is a repair of $\SD$ iff: $(i)$ $\D'$ is consistent with $\Sigma$; $(ii)$ for every $\alpha\in\D\setminus\D'$, $\{\alpha\}$ is not weakly consistent with $\SD$. Since condition $(i)$ can be checked in $\aczero$ w.r.t.\ data complexity (by Proposition~\ref{pro:consistency-ub}) and condition $(ii)$ can be checked in NL w.r.t.\ data complexity (by Theorem~\ref{thm:wc-lb-linear-ffk}), the thesis follows.
\qed

\bigskip

\noindent\textbf{Proof of Proposition~\ref{pro:algorithm-ars-correct}}:

The proof immediately follows from the fact that a set $\D'$ satisfying the three conditions of the algorithm exists if and only if there exists a repair of $\SD$ that does not entail $Q$.
\qed

\bigskip

\noindent\textbf{Proof of Theorem~\ref{thm:ar-ub-general-and-ffk}}:

First, observe that consistency of a database w.r.t.\ a set of dependencies can be decided in $\aczero$ w.r.t.\ data complexity, as well as the evaluation of a BUCQ over a database. Then, the proof immediately follows from Algorithm~\ref{alg:ars}, Proposition~\ref{pro:algorithm-ars-correct}, Proposition~\ref{pro:wc-ub-general},
and Theorem~\ref{thm:wc-ub-ffk}.
\qed

\bigskip

\noindent\textbf{Proof of Proposition~\ref{pro:algorithm-irs-correct}}:

It is immediate to see that, if for an image $M$ there exists no subset $\D'$ of $\D$ satisfying the conditions of the algorithm, then there exists no repair of $\SD$ that does not contain $M$, therefore $Q$ is \irs-entailed by $\SD$. Conversely, if, for every image $M$ of the query, such a subset $\D'$ exists, then $\D'$ belongs to at least a repair $\R$ of $\SD$ which does not contain $M$, and therefore $M$ is not contained in the intersection of all the repairs of $\SD$, thus no image of $Q$ is contained in the intersection of all the repairs, which implies that $Q$ is not \irs-entailed by $\SD$.
\qed

\bigskip

\noindent\textbf{Proof of Theorem~\ref{thm:ir-ub-general-and-ffk}}:

First, observe that consistency of a database w.r.t.\ a set of dependencies can be decided in $\aczero$ w.r.t.\ data complexity, as well as the evaluation of a BUCQ over a database. Then, the proof immediately follows from Algorithm~\ref{alg:irs}, Proposition~\ref{pro:algorithm-irs-correct}, Proposition~\ref{pro:wc-ub-general},
and Theorem~\ref{thm:wc-ub-ffk}.
\qed

\bigskip

\noindent\textbf{Proof of Theorem~\ref{thm:ar-ub-acyclic}}:

From the definition of \ars-entailment, it follows that a BUCQ $Q$ is not \ars-entailed by $\SD$ iff there exists a subset $\D'$ f $\D$ such that $(i)$ $\D'$ is a repair of $\SD$ and $(ii)$ $\D'\not\models Q$. Therefore, by Theorem~\ref{thm:rc-ub-acyclic} the thesis follows.
\qed

\bigskip

\noindent\textbf{Proof of Theorem~\ref{thm:ir-ub-acyclic}}:

First, observe that $Q$ is \irs-entailed by $\SD$ iff there exists an image of $Q$ in $\D$ such that $M\subseteq\intrep(\D)$, i.e.\ $M$ is contained in every repair of $\SD$.
Therefore, if $M_1,\ldots M_m$ are the images of $Q$ in $\D$, $Q$ is \irs-entailed by $\SD$ iff there exist $m$ repairs of $\SD$ $\R_1,\ldots,\R_m$ such that $M_i\not\subseteq\R_i$ for every $i$ such that $1\leq i\leq m$. Since the above number $m$ is bounded by $n^k$, where $n$ is the size of $\D$ and $k$ is the number of predicate atoms of $Q$, and by Theorem~\ref{thm:rc-ub-acyclic} repair checking in the case of acyclic dependencies is in $\aczero$ w.r.t.\ data complexity, the thesis follows.
\qed

\bigskip

\noindent\textbf{Proof of Theorem~\ref{thm:qe-ub-linear}}:

The proof follows immediately from Algorithm~\ref{alg:compute-repair-linear}: once computed (in PTIME) the only repair $\D'$ of $\SD$, the query is then evaluated over $\D'$ (which can be done in $\aczero$).
\qed

\bigskip

\noindent\textbf{Proof of Theorem~\ref{thm:qe-ub-linear-ffk}}:

Given the hypothesis, it is immediate to verify that $Q$ is \ars-entailed by $\SD$ iff there exists an image $M$ of $Q$ in $\D$ such that, for every fact $\alpha\in M$, $\{\alpha\}$ is weakly consistent with $\SD$. Since there are at most $n^k$ images of $Q$ in $\D$, where $n$ is the size of $\D$ and $k$ is the maximum number of predicate atoms in a BCQ of of $Q$, and since by Theorem~\ref{thm:wc-lb-linear-ffk} weak consistency can be checked in NL w.r.t.\ data complexity, the thesis follows.
\qed

\bigskip

\noindent\textbf{Proof of Theorem~\ref{thm:algorithm-acyclic-ffk-correct}}:

First, suppose $\SD$ \irs-entails $Q$. Then, there exists an image $M$ of $Q$ in $\D$ that belongs to all the repairs of $\SD$.
Now, suppose the algorithm returns false, and let $M$ and $\D'$ satisfy the three conditions (a), (b), (c) of the algorithm.
Since condition (b) holds, by Proposition~\ref{pro:wc-fc} $\D'$ is weakly consistent with $\SD$, thus $\D'\subseteq\R$ for some repair $\R$ of $\SD$. And, since by hypothesis $M$ is contained in each repair, we have that $\D'\cup M\subseteq\R$.
Since $\R$ is consistent with $\Sigma$ and contains $\D'\cup M$, it follows that $\D'\cup M$ is weakly consistent with $\Sigma$ too, thus by Proposition~\ref{pro:wc-fc} condition (c) is false, contradicting the above hypothesis. Consequently, the algorithm returns true.

Conversely, suppose the algorithm returns true and let $M$, $\D'$ satisfy conditions (a), (b), (c) of the algorithm.
Now suppose there exists a repair $\R$ of $\SD$ such that $M\not\subseteq\R$. This implies that $\R\cup M$ is not weakly consistent with $\SD$, therefore by Proposition~\ref{pro:wc-fc} $\fc(\R\cup M,\D,\Sigma)$ is inconsistent with $\Sigma$, hence by Proposition~\ref{pro:consistency} 
there exists a dependency in $\tau\in\Sigma$ and an instantiation $\sigma$ of $\body(\tau)$ in $\fc(\R\cup M,\D,\Sigma)$ such that the BUCQ $\sigma(\head(\tau))$ has no image in $\D$. Let $V$ be the set $\{\sigma(\alpha)\mid\alpha\in\predatoms(\body(\tau))\}$. Observe that $V$ contains at most $k$ facts, where $k$ is the maximum number of predicate atoms occurring in the body of a dependency in $\Sigma$.

Furthermore, since $\Sigma$ is acyclic, the presence of a fact in $\fc(\R\cup M,\D,\Sigma)$ depends on at most $k^h$ facts (where $h$ is the number of dependencies in $\Sigma$), therefore there exists a set of facts $\R'\subseteq\R$ such that $|\R'|\leq k^{h+1}$ and $\fc(\R'\cup M,\Sigma)$ contains $V$, causing $\fc(\R'\cup M,\Sigma)$ to be inconsistent with $\Sigma$.
But this implies that $\R'$ satisfies all the conditions of the set $\D'$ in the algorithm, which therefore returns false, thus contradicting the hypothesis. Hence, every repair of $\SD$ contains $M$, which implies that $Q$ is \irs-entailed by $\SD$.
\qed